\documentclass[a4paper,11pt]{article}
\pdfoutput=1
\usepackage{graphicx,array}
\usepackage{color}
\usepackage{latexsym}
\usepackage{amsthm}
\usepackage{amsmath}
\usepackage{amssymb}
\usepackage{empheq}

\usepackage{dsfont}
\usepackage{epsfig}
\usepackage{slashed}
\usepackage{bbold}
\usepackage{psfrag}
\usepackage[svgnames]{xcolor}
\PassOptionsToPackage{caption=false}{subfig}
\usepackage{subcaption}
\usepackage{xfrac}
\usepackage{multirow}
\usepackage{booktabs}

\usepackage[export]{adjustbox}
\usepackage{tikz}
\usepackage{tkz-euclide}
\usetikzlibrary{decorations.pathmorphing}	
\tikzset{
    v/.style={decorate, decoration={snake, segment length=3mm, amplitude=0.75mm}, draw},
    f/.style={draw=black, postaction={decorate},
        decoration={markings,mark=at position .6 with {\arrow[very thick]{latex}}}},
    fb/.style={draw=black, postaction={decorate},
        decoration={markings,mark=at position .4 with {\arrowreversed[very thick]{latex}}}},
    fnar/.style={draw=black},
    g/.style={decorate, draw=black,
        decoration={coil,amplitude=3pt, segment length=3.5pt}},
    s/.style={dashed,draw=black, postaction={decorate},
        decoration={markings,mark=at position .55 with {\arrow[very thick]{latex}}}},
    sb/.style={dashed,draw=black, postaction={decorate},
        decoration={markings,mark=at position .55 with {\arrowreversed[draw=black,very thick]{latex}}}},
    snar/.style={dashed,draw=black,line width =1.25pt},
}

\usepackage{cite}
\usepackage[normalem]{ulem}

\newcommand{\be}{\begin{equation}}
\newcommand{\ee}{\end{equation}}
\newcommand{\bea}{\begin{eqnarray}}
\newcommand{\eea}{\end{eqnarray}}

\usepackage{mathrsfs}

\newcommand{\nn}{\nonumber}
\def\({\left(}
\def\){\right)}

\usepackage{jheppub} 

\title{The see-saw portal at future Higgs Factories}

\author[a,b]{Daniele Barducci,}
\author[c]{Enrico Bertuzzo,}
\author[d,e]{Andrea Caputo,}
\author[f]{Pilar Hernandez,}
\author[b]{Barbara Mele}

\affiliation[a]{Universit\`a degli Studi di Roma la Sapienza, Piazzale Aldo Moro 5, 00185, Roma, Italy}
\affiliation[b]{INFN Section of Roma 1, Piazzale Aldo Moro 5, 00185, Roma, Italy}
\affiliation[c]{Instituto de Fisica, Universidade de Sao Paulo, C.P. 66.318, 05315-970 Sao Paulo, Brazil}
\affiliation[d]{School of Physics and Astronomy, Tel-Aviv University, Tel-Aviv 69978, Israel}
\affiliation[e]{Department of Particle Physics and Astrophysics,
Weizmann Institute of Science, Rehovot 7610001,Israel}
\affiliation[f]{Instituto de F\' isica Corpuscular - CSIC/Universidad de Valencia, Parc Cient\i fic de Paterna}

\emailAdd{daniele.barducci@roma1.infn.it}
\emailAdd{bertuzzo@if.usp.br}
\emailAdd{andrea.caputo@uv.es}
\emailAdd{m.pilar.hernandez@uv.es}
\emailAdd{barbara.mele@roma1.infn.it}

\abstract{
We consider an extension of the Standard Model with two right-handed singlet fermions with mass at the electroweak scale that induce neutrino masses, plus a generic new physics sector at a higher scale $\Lambda$. We focus on the effective operators of lowest dimension $d=5$, which induce new production and decay modes for the singlet fermions. We assess the sensitivity of future Higgs Factories, such as FCC-ee, CLIC-380, ILC and CEPC, to the coefficients of these operators for various center of mass energies. We show that future lepton colliders can test the cut-off of the theory up to $\Lambda \simeq 500 - 1000\;$TeV, surpassing the reach of future indirect measurements of the Higgs and $Z$ boson widths. We also comment on the possibility of determining the underlying model flavor structure should a New Physics signal be observed, and on the impact of higher dimensional $d=6$ operators on the experimental signatures.
}

\begin{document} 
\maketitle

\clearpage\newpage
\section{Introduction}
The discovery of neutrino masses and oscillations is one of the most striking evidences for the need of new physics (NP) beyond the Standard Model (SM). Arguably, the simplest extension of the SM consists in extending its field content with the right-handed (RH) counterparts of the left-handed SM neutrinos, $N$. At the renormalizable level this allows for Yukawa type interactions between the new states and the SM leptons, providing a Dirac mass term for the neutrinos. However, being  electroweak (EW) and color singlets, the new states can also have a Majorana mass terms, $M_N$. As it is well known, this allows to explain the lightness of the observed neutrino masses through a large hierarchy between the EW scale and the mass scale of the RH neutrinos
\be\label{eq:naive_see-saw}
m_\nu \propto y^2 \frac{v^2}{M_N},
\ee
where $y$ is the strength of the Yukawa interaction and $v$ the Higgs vacuum expectation value (VEV). The relation of Eq.~\eqref{eq:naive_see-saw} defines the see-saw mechanism~\cite{Minkowski:1977sc,Mohapatra:1979ia,Yanagida:1979as,GellMann:1980vs}. For a natural choice of the Yukawa interactions, $y={\mathcal O}(1)$, the lightness of neutrino masses requires RH neutrinos at around the Grand Unification scale. However see-saw models with EW-scale RH neutrinos have recently received increasing attention. On the one side they offer a compelling alternative for the generation of the matter-antimatter asymmetry via neutrino oscillations~\cite{Akhmedov:1998qx,Asaka:2005pn}, while on the other side they can be searched for at colliders and in beam-dump experiments~\cite{Keung:1983uu,Ferrari:2000sp,Graesser:2007pc,delAguila:2008cj,BhupalDev:2012zg,Helo:2013esa,Blondel:2014bra,Abada:2014cca,Cui:2014twa,Antusch:2015mia,Gago:2015vma,Antusch:2016vyf,Caputo:2016ojx,Das:2017zjc,Caputo:2017pit,Abada:2018sfh,Hernandez:2018cgc,Jones-Perez:2019plk}. 

At the renormalizable level these extra states can only be produced or decay via their mixing with the active neutrinos. This mixing, that controls their charged- and neutral-current interactions, is given by
\begin{eqnarray}
\theta \propto {y v\over M_N}.
\end{eqnarray}
For EW scale RH neutrinos the naive see-saw scaling of Eq.~\eqref{eq:naive_see-saw} requires a tiny value for the Yukawa coupling connecting the SM and the beyond the SM (BSM) sectors. This implies a tiny mixing of the RH neutrino, resulting in a small production cross-section and a small decay width. The latter can give rise to striking signatures, such as displaced decays.

Interestingly, both the properties of production via mixing and displaced decays of the RH neutrinos can be challenged. The naive see-saw scaling relation can be broken when more than one RH state is present by specific Yukawa and Majorana mass textures that ensure an approximate lepton number symmetry~\cite{Kersten:2007vk,Gavela:2009cd}. The mixing can be much larger than the one implied by the see-saw relation, thus modifying the lifetime of the BSM states. RH neutrinos can then feature a prompt, displaced or detector stable behaviour.

These predictions can also be altered by the presence of additional NP states at a scale $\Lambda \gg v, M$. At low energy their effects can be described in the language of effective field theories (EFT) by a tower of higher dimensional operators ${\cal O}^d \Lambda^{4-d}$ with $d>4$, built out from the SM and RH neutrinos fields: the $\nu$SMEFT. At the lowest  dimension, $d=5$, two new operators intervene to induce new RH neutrinos production modes: an operator triggering a new Higgs decay channel into a pair of RH neutrinos and a dipole operator connecting the RH neutrino tensor current with the hypercharge gauge boson~\cite{Graesser:2007yj,Aparici:2009fh}. At $d=6$ many more operators are present~\cite{Graesser:2007pc,delAguila:2008ir,Liao:2016qyd}, which can induce  new production as well as new decay channels.

Various theoretical studies have investigated the signatures at the Large Hadron Collider (LHC) of a subset of these higher dimensional operators, see {\emph{e.g.}}~\cite{Graesser:2007yj,Graesser:2007pc,Aparici:2009fh,Caputo:2017pit,Butterworth:2019iff,Alcaide:2019pnf,deVries:2020qns}. The search for EW scale RH neutrinos is however one the primary goal of future $e^+ e^-$ colliders, thanks to the clean detector environment and the tipically lower SM backgrounds with respect to an hadronic machine, which can help to overcome the generally small production cross-sections of SM singlet states. Various future prototypes has been designed for the post LHC era: both circular colliders, as the Future Circular Collider~\cite{Gomez-Ceballos:2013zzn,Abada:2019zxq,Abada:2019lih,Blondel:2019yqr} (FCC-ee) and the Compact electron-positron collider~\cite{CEPCStudyGroup:2018rmc,CEPCStudyGroup:2018ghi} (CEPC), and linear ones, such as the International Linear Collider~\cite{Behnke:2013xla,Baer:2013cma,Bambade:2019fyw} (ILC) and the Compact Linear Collider~\cite{deBlas:2018mhx,Roloff:2018dqu} (CLIC). 
It is then the purpose of this paper to investigate the phenomenology of the  $\nu$SMEFT at these future machines and study their sensitivity on the $d>4$ operators inducing new RH neutrinos production and decay modes in all the possible regimes of the $N$ decay lifetimes, for RH masses in the range 1\,--\,60 GeV.

The paper is organized as follows. In Sec.~\ref{sec:th} we introduce our notation and describe the active-sterile mixing formalism while in Sec.~\ref{sec:colliders} we discuss the properties of the various $e^+ e^-$ colliders under analysis. In Sec.~\ref{sec:ONH} and in Sec.~\ref{sec:ONB} we describe the analysis details and show the projected reach of the collider prototypes on the  $d=5$ operators involving RH neutrino fields, while in Sec.~\ref{sec:d6} we discuss the possible impact of $d=6$ operators inducing extra $N$ production and decay modes. We then conclude in Sec.~\ref{sec:conc}.

\section{Theoretical framework}\label{sec:th}

We work in the framework of the $\nu$SMEFT, which is described by the following Lagrangian
\be\label{eq:lag_nusmeft}
{\cal L} = {\cal L}_{\rm SM}  + \bar N \slashed \partial N - \bar L_L Y_\nu \tilde H N - \frac{1}{2} M_{N} \bar N^c N + \sum_{n>4} \frac{{\cal O}^n}{\Lambda^{n-4}} + h.c.
\ee
where $N$ is vector describing ${\cal N}$ flavors of gauge singlet RH neutrino fields with $N^c = C \bar N^T$ and $C= i \gamma^2 \gamma^0$,  $L$ is the SM lepton doublet, $Y_\nu$ is the $3\times {\cal N}$ Yukawa matrix of the neutrino sector with $\tilde H = i \sigma^2 H^*$, $M_{N}$ is a ${\cal N}\times {\cal N}$ Majorana mass matrix for the RH neutrino fields and ${\cal O}^n$ the Lorentz and gauge invariant operators built out from the SM and the RH neutrino fields. The $\nu$SMEFT has been constructed up to $d=7$ in~\cite{Graesser:2007yj,Graesser:2007pc,delAguila:2008ir,Aparici:2009fh,Liao:2016qyd}.
At dimension five only three operators exist
\begin{align}\begin{aligned}
      & {\cal O}_{W} = \alpha_W(\bar L^c \tilde H^*)( \tilde H^\dag L ) \ ,  \\
      & {\cal O}_{NH} = \alpha_{NH}(\bar N^c N )(H^\dag H ) \ , \\
&  {\cal O}_{NB} = \alpha_{NB}\bar N^c \sigma^{\mu\nu} N B_{\mu\nu}  \ ,
\end{aligned}
\end{align}
where $\alpha_W$ and $\alpha_{NH}$ are symmetric $3\times 3$ and ${\cal N}\times {\cal N}$ matrices in flavor space respectively, $\alpha_{NB}$ is an antisymmetric ${\cal N}\times {\cal N}$ matrix in flavor space, $\sigma^{\mu\nu}=i/2 [\gamma^\mu,\gamma^\nu]$ and $B_{\mu\nu}$ is the SM hypercharge field strength tensor.
The first operator is the well known Weinberg operator~\cite{Weinberg:1979sa} responsible for generating a Majorana mass for the SM neutrinos. The operator ${\cal O}_{NH}$ induces new interactions between the Higgs field and the RH neutrinos and adds an extra contribution to the RH neutrinos Majorana mass matrix, while ${\cal O}_{NB}$ is a dipole type operator connecting the RH neutrino tensor current to the hypercharge gauge boson. The operator coefficients $\alpha_W$
and $\alpha_{NH}$ can be ${\cal O}(1)$, while $\alpha_{NB}$ is necessarily ${\cal O}(1/16\pi^2)$ since it can be only generated at loop level in a weakly coupled ultraviolet completion of the effective Lagrangian of Eq.~\eqref{eq:lag_nusmeft}.

\subsection{Neutrino mixing formalism}\label{sec:mixing-formalism}

Without loss of generality it is possible to go from Eq.~\eqref{eq:lag_nusmeft} to a basis where the matrix $M_{N}$ and the charged lepton mass matrix are diagonal with non negative entries. Working at $d=5$, the operator ${\cal O}_{NH}$ contributes to the neutrino mass matrix.
By defining $n = (\nu_L, N^c)$ and using $\langle H \rangle=174\;$GeV, the mass Lagrangian in the neutrino sector can be written as

\be\label{eq:nu_mass_matrix}
{\cal L}_{\rm mass} = -\frac{1}{2}\bar n^c {\cal M} n + h.c.  = -\frac{1}{2}\bar n^c \left(
\begin{array}{ccc}
- 2\alpha_{W} \, \frac{ v^2}{\Lambda}  & &Y_\nu v  \\  
 & &\\
Y_\nu^T v  & & M_N - 2 \alpha_{NH} \, \frac{ v^2}{\Lambda}
\end{array}
\right) n   + h.c. \ ,
\ee
where the $\nu_L - \nu_L$ block receives a contribution only from the $d=5$ Weinberg operator while the $N - N$ one has both $d=4$ and $d=5$ contributions. This mass matrix can be perturbatively diagonalized in the regime in which the entries of the $\nu_L - N$ block are smaller than the ones in the $N - N$ one.  
 For our purposes we assume that the see-saw contribution to the active neutrino masses dominates over the other ones. Under this approximation we obtain
\be\label{eq:nu_mass_def}
m_\nu \simeq v^2 Y_\nu \frac{1}{M_N} Y_\nu^T = U^* m_\nu^{(d)} U^\dag \ ,
\ee
where $m_\nu^{(d)}$ is diagonal with non negative entries and $U$ is the Pontecorvo-Maki-Nakagawa-Sakata (PMNS) matrix~\cite{Pontecorvo:1957qd,Maki:1962mu}. From Eq.~\eqref{eq:nu_mass_def} one can obtain
\be
Y_\nu \simeq \frac{1}{v}U^* \sqrt{\mu}\sqrt{M_N} \ ,
\ee
where the $3\times {\cal N}$ matrix $\sqrt \mu$ satisfies $\sqrt \mu \sqrt \mu^T = m_\nu^{(d)}$ and $\sqrt{\mu}$ and $\sqrt{M_N}$ indicate, respectively, $\mu^{1/2}$ and $M_N^{1/2}$. The usefulness of this parametrization is that it allows to write in a compact way the expressions for the various matrices involved. We now restrict our analysis to the case of two RH neutrinos, thus fixing ${\cal N}=2$.
Without loss of generality the matrix $\sqrt{\mu}$ can be written using the so-called Casas-Ibarra parametrization~\cite{Casas:2001sr} as
\be
\sqrt{\mu} = \sqrt{m}\, \mathcal{R}\ ,
\ee
where $\sqrt{m}$ is a $3\times 2$ matrix containing the physical neutrino masses $m_i$, while $\mathcal{R}$ is a complex orthogonal $2\times 2$ matrix, ${\cal R}^T {\cal R}=\mathbb 1$. With two RH neutrinos one has $m_{\nu_1}=0$ and $m_{\nu_3}> m_{\nu_2}$ in the normal hierarchy (NH) case, while  $m_{\nu_3}=0$ and $m_{\nu_2}> m_{\nu_1}$ in the inverted hierarchy (IH) one~\footnote{For the NH we take $m_{\nu_2}=8.6 \times 10^{-3}\;$eV and $m_{\nu_3}=5.1\times 10^{-2}\;$eV while for the IH we take $m_{\nu_1}=4.9 \times 10^{-2}\;$eV and $m_{\nu_2}=5.0\times 10^{-2}\;$eV.}. More in detail, for NH and inverted IH we have
\be
\sqrt{m_{NH}} = \begin{pmatrix}
0 & 0 \\
0 & \sqrt{m_2} \\
\sqrt{m_3} & 0
\end{pmatrix}\ , ~~~ 
\sqrt{m_{IH}} = \begin{pmatrix}
0 & \sqrt{m_1} \\
 \sqrt{m_2}  & 0\\
0& 0
\end{pmatrix} \ ,
\ee
while we parametrize the orthogonal matrix $\mathcal{R}$ in terms of the complex angle $z = \beta + i \gamma$ as
\be\label{eq:Rmatrix}
\mathcal{R} = 
\begin{pmatrix}
\cos z & \pm \sin z \\
- \sin z & \pm \cos z
\end{pmatrix}\ .
\ee
For both hierarchies we can thus write
\be
Y_\nu \simeq \frac{1}{v}U^* \sqrt{m}{\cal R} \sqrt{M_N} \ ,
\ee 
where $m=m_{\rm NH}$ or $m_{\rm IH}$, and obtain a compact expression for the active-sterile mixing angle
\be
\theta_{\nu N} \simeq - U^* \sqrt{m} {\cal R} \frac{1}{\sqrt{M_N}} \ .
\ee
It's crucial that the angle $z$ can be taken in general as a complex number. In fact, in the limit in which $z$ is a real number, by taking $U$ and ${\cal R}$ with entries of order unity and by assuming an equal value for the diagonal entries of the Majorana mass term for the two RH neutrino  $m_{N_1}=m_{N_2}=m_N$, one obtains~\footnote{We have assumed NH and fixed $m_\nu = m_{\nu_3}$. The expression holds also for the IH case modulo order one factors.}
\be\label{eq:Ynu_realz}
Y_\nu \sim \frac{\sqrt{m_N}m_\nu}{v} \sim 4 \times 10^{-8} \left(\frac{m_N}{1\;{\rm GeV}} \right)^{1/2} \ .
\ee
This ``naive see-saw scaling'' relation is drastically modified by the imaginary part of $z$, that gives an exponential enhancement. In the limit $\gamma \gg 1$
\be
\mathcal{R} \simeq \frac{e^{\gamma- i \beta}}{2}
\begin{pmatrix}
1 & \pm i \\
- i & \pm 1
\end{pmatrix}\ ,
\ee
and the relation of Eq.~\eqref{eq:Ynu_realz} is modified to
\be
Y_\nu \sim  2 \times 10^{-8}  e^{\gamma -  i \beta}\left(\frac{m_N}{1\;{\rm GeV}} \right)^{1/2} \ .
\ee
The same enhancement is inherited by the active-sterile mixing, that now reads
\be\label{eq:increase}
 \theta_{\alpha i}  \equiv \left(\theta_{\nu N} \right)_{\alpha i} \sim  7.2 \times 10^{-6} \, e^{\gamma-i \beta} \, \left(\frac{1~\mathrm{GeV}}{m_N} \right)^{1/2}\ .
\ee
In the previous expression we have $\alpha=e,\mu,\tau$ and $i=1,2$. As anticipated in the Introduction and as we will see more in detail below, this deviation from the naive see-saw scaling has a crucial impact on the RH neutrinos phenomenology, especially for what concerns their decay width and consequently their lifetime, with huge implications for search strategies at future colliders.

\subsection{Heavy neutrinos decay modes}\label{sec:decay}

In the mass range of our interest and at the renormalizable level the RH neutrinos can only decay through charged- and neutral-currents via an off-shell $W$ or $Z$ boson~\footnote{A decay into an off-shell Higgs boson is generally suppressed by the smallness of the SM Yukawa couplings.}. The RH neutrino decay mode is thus completely fixed once the $W$ and $Z$ decay channels are specified. The various final states from $N$ decay are reported in Tab.~\ref{tab:N_decay} where $\alpha,\beta,i$ and $j$ are flavor indices and, for simplicity, we do not specify neither the charge of the charged lepton $\ell^\prime=e, \mu, \tau$ nor the nature of the (anti)neutrino. In this table we have grouped together the three decay modes giving rise to the $\nu\ell^\prime\ell^\prime$ final state since, given that both the $W$ and $Z$ boson will be non resonant for the RH neutrinos mass range of our interest, these processes will not be distinguishable. Moreover, the $\nu_\alpha \ell^\prime_\beta \ell^\prime_\beta$ process with $\alpha=\beta$ receives contributions from both neutral- and charged-current interactions, which interfere among themselves.

\begin{table}[t!]
\begin{center}
\begin{tabular}{c|c|c}
    Final state & Channel & Mediator \\
    \hline\hline
    $\ell^\prime q \bar q$ & $\ell^\prime_\alpha q_i \bar q_j$ & $W$ \\
    \hline
    $\nu q \bar q$ & $\nu_\alpha q_i \bar q_j$ &  $Z$ \\
    \hline
    \multirow{3}{*}{$\nu \ell^\prime \ell^\prime$}& $\ell^\prime_\alpha \ell^\prime_\beta \nu_\beta$, $\alpha \ne \beta$ &   $W$\\
    &$\nu_\alpha \ell^\prime_\beta \ell^\prime_\beta$, $\alpha \ne \beta$ &  $Z$\\
    &$\nu_\alpha \ell^\prime_\beta \ell^\prime_\beta$, $\alpha = \beta$ &  $W$ and $Z$\\
    \hline
    $\nu\nu\nu$ & $\nu_\alpha \nu_\beta \nu_\beta$ &  $Z$ \\
\end{tabular}
\end{center}
\caption{Possible decay channels for the RH neutrino $N$. Here $\alpha,\beta,i$ and $j$ are flavor indices and we do not specify the charge of the charged lepton $\ell^\prime=e, \mu, \tau$ nor the nature of the (anti)neutrino.}\label{tab:N_decay}
\end{table}

\begin{figure}[t!]
\begin{center}
\includegraphics[width=0.48\textwidth]{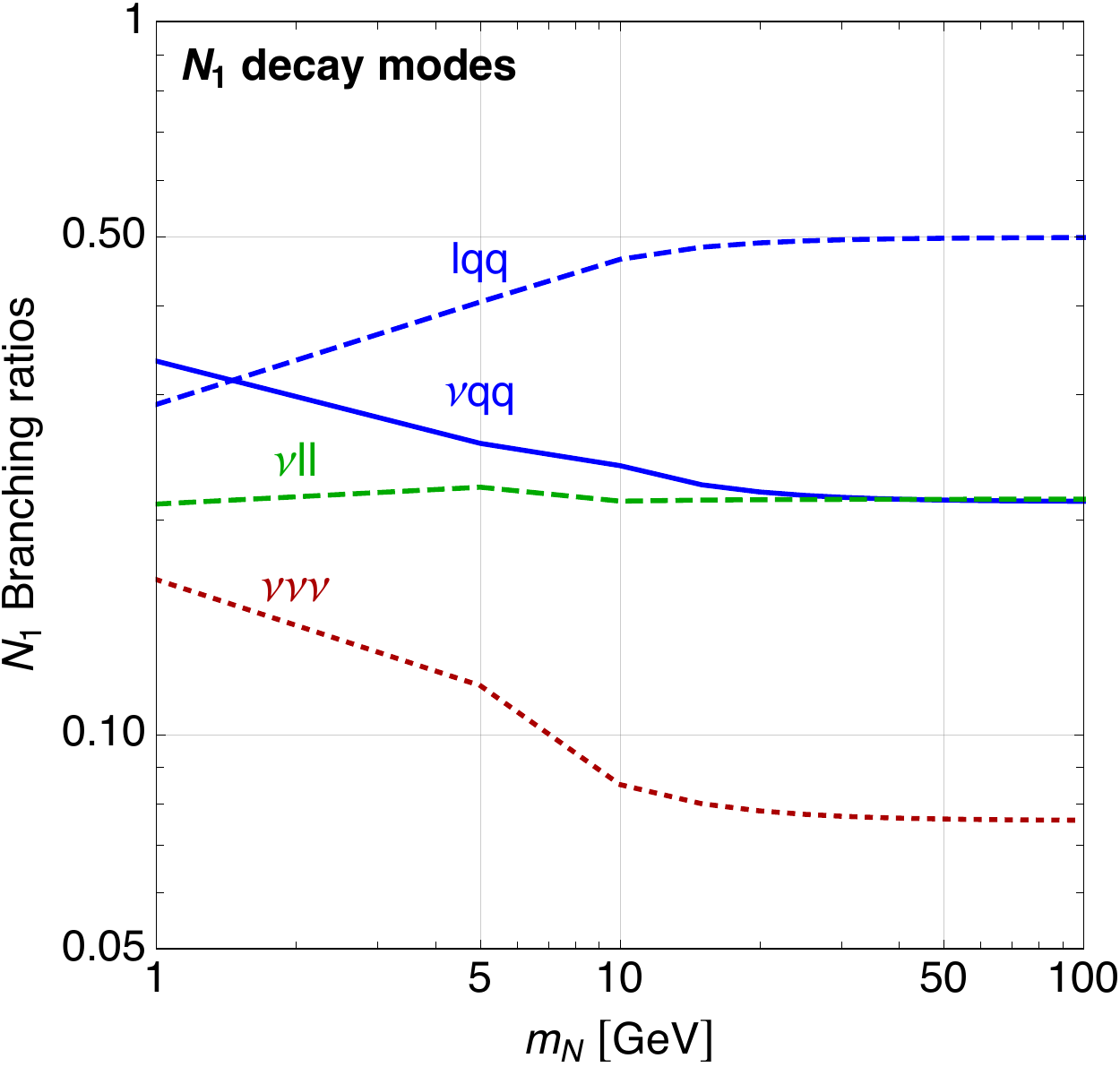}\hfill
\includegraphics[width=0.46\textwidth]{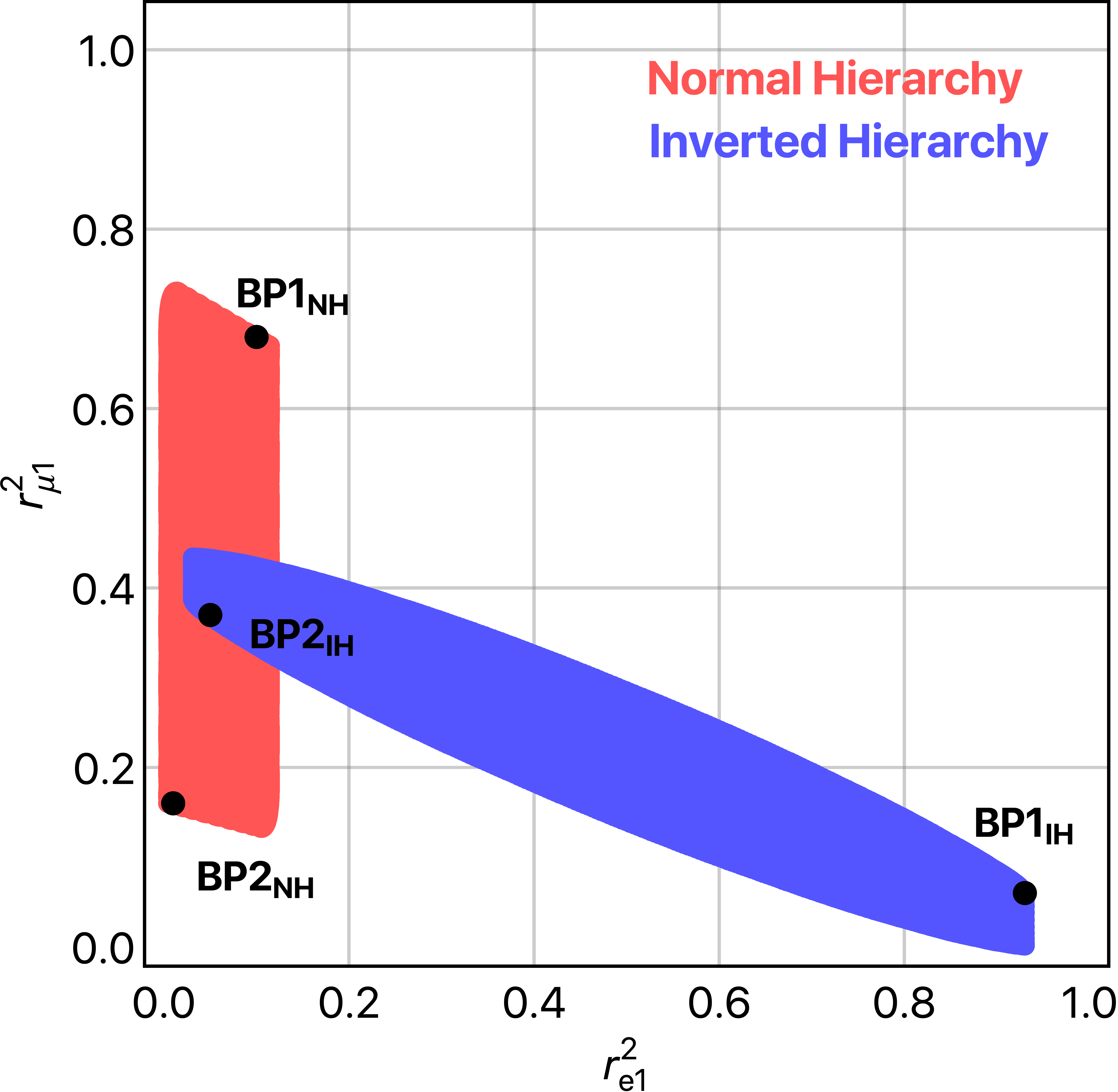}
\caption{\small {\emph{Left:}} Decay modes of the lightest RH neutrino $N_1$. We have fixed $\delta=\phi_1=\beta=\gamma=0$. {\emph{Right:}} Allowed ranges in the $r_{e1}^2-r_{\mu1}^2$ plane for $\gamma \gtrsim 1$, obtained by varying the PMNS phases $\delta$ and $\phi$ in the range $[0,2\pi]$ with  $r_{\tau1}=1-r_{e1}^2-r_{\mu1}^2$. Also shown are the benchmark points used in the analysis.}
\label{fig:NR_decay}
\end{center}
\end{figure}

For computing the partial widths into the final state of Tab.~\ref{tab:N_decay} we use the results of~\cite{Bondarenko:2018ptm}. For $m_N \gg \Lambda_{{\rm QCD}}$, the decay rates involving quark pairs are physical quantities, otherwise decays into hadrons should instead be considered. Following again~\cite{Bondarenko:2018ptm} we have implemented three-loop QCD corrections through which the full hadronic width can be computed from the decay width into free quarks. Altogether the effect on the total width is found to be around 30\% for $m_N \simeq 1\;$GeV, decreasing down to 10\% for $m_N \simeq 5\;$GeV. By fixing the phases of the PMNS matrix $\delta=\phi_1=0$~\footnote{We remind the reader that with $\mathcal{N}=2$ RH states only two phases are present in the PMNS matrix. We denote by $\delta$ the so-called Dirac phase, and by $\phi_1$ the so-called Majorana phase.} and also $\gamma=\beta=0$ we obtain the branching ratios (BRs) shown in the left panel of Fig.~\ref{fig:NR_decay} for the case of the lightest RH neutrino $N_1$, while similar rates are obtained for $N_2$.

The lepton flavor composition of the final states depends on the active-sterile mixing matrix, which in turn depends on {\emph{i)}} the  hierarchy and the squared mass differences of the active neutrinos, {\emph{ii)}}  the phases of the PMNS matrix $\delta$ and $\phi_1$, and {\emph{iii)}}  the $\beta$ and $\gamma$ parameters entering in the Casas-Ibarra parametrization of Eq.~\eqref{eq:Rmatrix}. Analytical approximations for the various mixings can be derived when $\gamma \gtrsim 1$, see {\emph{e.g.}}~\cite{Hernandez:2016kel}. In this regime the Casas-Ibarra parameter $\beta$ has a little impact on the normalized squared mixing
\be
r^2_{\alpha i} = \frac{|\theta_{\alpha i}|^2}{\bar U^2_i }, \qquad \bar U^2_i = \sum_{\alpha=e,\mu,\tau} |\theta_{\alpha i} |^2, \qquad \alpha=e,\mu,\tau,\quad i=1,2
\ee
which is mainly determined by the PMSN phases $\delta$ and $\phi_1$. By varying them between [0,$2\pi$] we obtain for $r^2_{\alpha 1}$ the allowed ranges shown in the right panel of Fig.~\ref{fig:NR_decay}. The red and blue regions correspond to NH and IH, respectively, and $r^2_{\tau1}=1-r_{e1}^2-r_{\mu1}^2$, see also~\cite{Caputo:2017pit}. We also show the benchmark points that will be used in the following analysis.

\subsection{Final states from pair produced heavy neutrinos}

Both the operators ${\cal O}_{NH}$ and ${\cal O}_{NB}$ will mediate the production of a pair of RH neutrinos, which in turn will decay producing a six-body final state. These final states can be categorized into fully-leptonic, fully-hadronic, semi-leptonic and invisible channels and are reported in Tab.~\ref{tab:N1N1_decay}. Here we differentiate between $\ell=e,\mu$ and $\tau$, since the latter particle decays within the detector thus producing a different and more complex final state. With an abuse of notation we nevertheless use the nomenclature fully-leptonic and semi-leptonic also for the final states involving $\tau$ leptons. Particularly interesting are the final states that can present a pair of same-sign (SS) leptons $\ell=e,\mu$, a signature which generally has a low SM background. Final states with $\ge 3\ell$ will clearly have a SS pair. However, also the final state with exactly $2\ell$ can produce a SS signature, due to the Majorana nature of the RH neutrinos. The branching ratios of $N_1$ and $N_2$ into the various final  states of Tab.~\ref{tab:N1N1_decay} will depend upon the choice of the normalized squared mixings $r^2_{\alpha i}$ which, as explained above, depend on the PMNS parameter $\delta$ and $\phi_1$ and can span the ranges illustrated on the right panel of Fig.~\ref{fig:NR_decay}. 

For what concerns the detection and reconstruction of the final state, $e$ and $\mu$ can be considered in first approximation on the same footage, while one needs to distinguish them with respect to $\tau$ leptons. We then choose to perform our analysis and illustrate our results for some representative points in the allowed range for the $r^2_{\alpha i}$ shown in Fig.~\ref{fig:NR_decay} for both the NH and IH cases. In particular for each mass hierarchy we choose two points, one with a large and one with a small mixing with the third generation leptons.  After having fixed $r^2_{\tau i}$ we choose to maximize the mixing with the electron, which does not affect the reconstruction of the final state, in the approximation of  similar $e$ and $\mu$ detection efficiencies, but does intervene in the production of RH neutrinos via mixing, $e^+ e^- \to \nu N$. The reason why we choose to maximize $r^2_{e\alpha}$ is because, conservatively, we want to analyze a configuration with the largest possible production cross-section via mixing, to see whether the additional production modes arising from the ${\cal O}_{NH}$ and ${\cal O}_{NB}$ operators can still dominate over it. More concretely, we choose the following two benchmark points for the NH case
\begin{align}\label{eq:BP_NH}
& {\bf BP1_{NH}:}\quad r^2_{e4} : r^2_{\mu4} : r^2_{\tau 4}  =  0.10 : 0.68  : 0.22  \\ 
\nn & {\bf BP2_{NH}:}\quad r^2_{e4} : r^2_{\mu4} : r^2_{\tau 4}  =  0.01 : 0.16  : 0.83 \ ,
\end{align}
which, in the $\gamma > 1$ and degenerate mass limit, can be realized simultaneously for both $N_1$ and $N_2$, which therefore have similar BRs. The corresponding $NN$ decay rates are reported in Tab.~\ref{tab:BP12_NH}. Similarly,  we choose the following benchmark points for the IH case
\begin{align}\label{eq:BP_IH}
& {\bf BP1_{IH}:}\quad r^2_{e4} : r^2_{\mu4} : r^2_{\tau 4}  =  0.93 : 0.06  :  0.01  \\ 
\nn & {\bf BP2_{IH}:}\quad r^2_{e4} : r^2_{\mu4} : r^2_{\tau 4}  =  0.05 : 0.37  : 0.58 \ ,
\end{align}
whose BRs are reported in Tab.~\ref{tab:BP12_IH}.

\begin{table}[t]
\begin{center}
\scalebox{0.75}{
\begin{tabular}[t]{c|c|c}
       & Channel & SS \\
\hline
\hline
   \multirow{2}{*}{Fully-leptonic}  & 4$\ell$ $\slashed E_T$ & {\Large \checkmark}\\
  						 & $2\ell$ $\slashed E_T$& \\  
\hline						 
\multirow{5}{*}{Semi-leptonic}   & $3\ell$ $2q$  $\slashed E_T$  &{\Large \checkmark} \\
    & $2\ell$ $4q$ & {\Large \checkmark}\\  
  & $2\ell$ $2q$ $\slashed E_T$ &\\  
  &  $\ell$ $4q$ $\slashed E_T$ & \\
   & $\ell$ $2q$  $\slashed E_T$ &\\
\hline 
\multirow{2}{*}{Fully-hadronic}     & $4q$ $\slashed E_T$ &\\
  & $2q$ $\slashed E_T$& \\
\hline  
Invisible  & $\slashed E_T$& \\
\end{tabular}}
\quad
\scalebox{0.8}{
\begin{tabular}[t]{c|c|c}
      & Channel  & SS\\
\hline\hline
\multirow{6}{*}{Fully-leptonic}  & $3\ell$ $\tau$ $\slashed E_T$  & {\Large \checkmark}\\
    & $2\ell$ 2$\tau $ $\slashed E_T$& \\  
    & $\ell$ $\tau$  $\slashed E_T$ &  \\  
    & $\ell$ 3$\tau $  $\slashed E_T$& \\   
    & 4$\tau $ $\slashed E_T$& \\  
    & 2$\tau $ $\slashed E_T$& \\
 \hline           
\multirow{4}{*}{Semi-leptonic}     & $2\ell$ $\tau$ $2q$ $\slashed E_T$   &\\
  &  $\ell$ 2$\tau $  $2q$ $\slashed E_T$ &\\
    & $\ell$ $\tau$ $4q$   &\\  
   & $\ell$ $\tau$ $2q$  $\slashed E_T$  & \\   
 \end{tabular}
}
\quad
\scalebox{0.8}{
\begin{tabular}[t]{c|c|c}
      & Channel  & SS\\
 \hline\hline           
\multirow{5}{*}{Semi-leptonic}     & 3$\tau $ $2q$  $\slashed E_T$   &\\
  & 2$\tau $ $4q$ &\\
  & 2$\tau $ $2q$  $\slashed E_T$ &\\    
  & $\tau$ $2q$ $\slashed E_T$ &  \\
  & $\tau$ $4q$ $\slashed E_T$ &  \\  
\end{tabular}
}
\quad
\end{center}
\caption{Possible final states from the decay of pair produced RH neutrinos. The checkmarks correspond to channels that can produce a SS leptons signal. The leftmost and rightmost tables contains final states without $\tau$ leptons and light leptons, $\ell$, respectively. }\label{tab:N1N1_decay}
\end{table}

\begin{table}[t!]
\begin{center}
\scalebox{0.75}{
\begin{tabular}{c|c|c||}
BR & Channel  & SS\\
\hline
 0.16 &  2$\ell$   4$q$  &  {\Large \checkmark}\\
 0.09 &  $\ell$ 4$q$    $\slashed E_T$  & \\
 0.05 &  4$q$   $\slashed E_T$  & \\
 0.05 &  2$\ell$ $\tau$   2$q$   $\slashed E_T$    & \\
 0.04 &  3$\ell$ 2$q$      $\slashed E_T$  &  {\Large \checkmark}\\
 0.03 &  $\ell$ 4$q$   $\tau$  & \\
 0.03 &  $\ell$ 2$q$    $\slashed E_T$  & \\
 0.02 &  2$\ell$   2$q$   $\slashed E_T$  & \\
 0.02 &  $\ell$ $\tau$ 2$q$    $\slashed E_T$    & \\
 0.02 &  $\tau$ 4$q$   $\slashed E_T$    & \\
 0.02 &  2$q$   $\slashed E_T$  & \\
 0.01 &  $\ell$  2$\tau $ 2$q$      $\slashed E_T$  & \\
 0.01 &  4$\ell$   $\slashed E_T$  &  {\Large \checkmark}\\
\end{tabular}}
\scalebox{0.75}{
\begin{tabular}{c|c | c}
BR & Channel & SS  \\
\hline
 0.01 &  3$\ell$ $\tau$  $\slashed E_T$    & \\
 0.01 &  2$\ell$   2$\tau $   $\slashed E_T$  & \\
 0.01 &  2$\ell$   $\slashed E_T$  & \\
 0.01 &  2$\tau $   4$q$  & \\
 0.01 & $\ell$ $\tau$ $\slashed E_T$    & \\
 0.01 &  $\tau$ 2$q$   $\slashed E_T$    & \\
 0.01 &  $\slashed E_T$  & \\
 0. &  2$\tau $ 2$q$      $\slashed E_T$  & \\
 0. &  $\ell$ 3$\tau $    $\slashed E_T$  & \\
 0. &  3$\tau $ 2$q$      $\slashed E_T$  & \\
 0. &  2$\tau $   $\slashed E_T$  & \\
 0. &  4$\tau $   $\slashed E_T$  & \\
 &\\
\end{tabular}}
\hskip 8mm
\scalebox{0.75}{\begin{tabular}{c|c|c||}
BR & Channel  & SS\\
\hline
 0.13 &  2$\tau $   4$q$  &  \\
 0.09 &  $\tau$ 4$q$   $\slashed E_T$    & \\
 0.06 &  4$q$   $\slashed E_T$  & \\
 0.06 &   $\ell$ 2$\tau $ 2$q$       $\slashed E_T$  & \\
 0.04 & $\ell$ $\tau$ 2$q$    $\slashed E_T$    & \\
 0.03 &  $\ell$ $\tau$ 4$q$       & \\
 0.03 & $\tau$ 2$q$   $\slashed E_T$    & \\
 0.02 &  $\ell$ 4$q$    $\slashed E_T$  & \\
 0.02 &  2$\ell$   2$\tau $   $\slashed E_T$    & \\
 0.02 &  2$\ell$ $\tau$  2$q$   $\slashed E_T$  & \\
 0.02 &  2$q$   $\slashed E_T$  & \\
 0.01 &   3$\tau $ 2$q$     $\slashed E_T$  & \\
 0.01 & $\ell$ $\tau$ $\slashed E_T$    & \\
\end{tabular}}
\scalebox{0.75}{\begin{tabular}{c|c | c}
BR & Channel & SS  \\
\hline
 0.01 &  2$\tau $  2$q$     $\slashed E_T$  &{\Large \checkmark} \\
 0.01 &  2$\ell$   4$q$  & \\
 0.01 &  $\ell$ 2$q$    $\slashed E_T$  & \\
 0.01 &  $\slashed E_T$  & \\
 0.01 &  2$\ell$   2$q$   $\slashed E_T$  & \\
 0.01 &  $\ell$ 3$\tau $    $\slashed E_T$  & \\
 0. &  3$\ell$  $\tau$ $\slashed E_T$    & \\
 0. &  2$\tau $   $\slashed E_T$  & \\
 0. &  3$\ell$ 2$q$      $\slashed E_T$  & {\Large \checkmark} \\
 0. &  2$\ell$   $\slashed E_T$  & \\
 0. &  4$\tau $   $\slashed E_T$  & \\
 0. &  4$\ell$  $\slashed E_T$  & {\Large \checkmark} \\
 &\\ 
\end{tabular}}
\end{center}
\caption{Decay rates from $N N$ production for $\bf{BP1_{NH}}$ (left) and $\bf{BP2_{NH}}$ (right). The rates are obtained by summing on all charges and flavor configurations. Here $\ell=e,\mu$. The checkmarks correspond to channels that can produce a SS lepton signal.}\label{tab:BP12_NH} 
\end{table}

\begin{table}[t!]
\begin{center}
\scalebox{0.75}{
\begin{tabular}{c|c|c||}
BR & Channel  & SS\\
\hline
 0.24 &  2$\ell$   4$q$  &  {\Large \checkmark}\\
 0.11 &  $\ell$ 4$q$    $\slashed E_T$  & \\
 0.07 & 3$\ell$ 2$q$ $\slashed E_T$  &  {\Large \checkmark} \\
 0.05 & 4$q$ $\slashed E_T$ & \\
 0.04 & $\ell$ 2$q$ $\slashed E_T$ & \\
 0.04 & 2$\ell$ $\tau$ 2$q$ $\slashed E_T$ & \\
 0.03 & 2$\ell$ 2$q$ $\slashed E_T$ & \\
 0.02 & 4$\ell$ $\slashed E_T$ &  {\Large \checkmark} \\
 0.02 & 2$q$ $\slashed E_T$ & \\
  0.02 & $\ell$ $\tau$ 2$q$ $\slashed E_T$ & \\
  0.01 & 2$\ell$ $\slashed E_T$ & \\
  0.01 & 3$\ell$ $\tau$ $\slashed E_T$ & \\
  0.01 & $\slashed E_T$ & \\
\end{tabular}}
\scalebox{0.75}{
\begin{tabular}{c|c | c}
BR & Channel & SS  \\
\hline
 0.01 &  $\ell$ $\tau$  $\slashed E_T$    & \\
 0.01 &  2$\ell$ 2$\tau$  $\slashed E_T$    & \\
 0. &  $\ell$ $\tau$  4$q$    & \\
 0. & $\tau$ 4$q$ $\slashed E_T$    & \\
  0. & $\ell$ 2$\tau$ 2$q$ $\slashed E_T$    & \\
 0. & $\tau$ 2$q$ $\slashed E_T$    & \\
  0. & 2$\tau$ 2$q$ $\slashed E_T$    & \\
    0. & 2$\tau$ $\slashed E_T$    & \\
     0. & $\ell$ $3\tau$ $\slashed E_T$    & \\   
      0. & 2$\tau$ 4$q$     & \\       
      0. &  $3\tau$ 2$q$ $\slashed E_T$    & \\        
       0. &  $4\tau$ $\slashed E_T$    & \\          
  & \\
\end{tabular}}
\hskip 8mm
\scalebox{0.75}{\begin{tabular}{c|c|c||}
BR & Channel  & SS\\
\hline
 0.06 &     2$\tau$ 4$q$    & \\        
  0.06 &     $\tau$ 4$q$  $\slashed E_T$    & \\    
    0.06 &     4$q$ $\slashed E_T$      &   \\      
        0.05 &     $\ell$ $\tau$ 4$q$        & \\ 
        0.05 & $\ell$  4$q$ $\slashed E_T$& \\
       0.05 & 2$\ell$ 4$q$ & {\Large \checkmark} \\
0.04 & 2$\ell$ 2$\tau$ 2$q$          $\slashed E_T$    & \\ 
0.04 & $\ell$ 2$\tau$ 2$q$ $\slashed E_T$    & \\ 
  0.03 & $\ell$ $\tau$ 2$q$ $\slashed E_T$    & \\  
  0.02 & 2$q$ $\tau$ $\slashed E_T$    & \\  
  0.02 &  2$q$ $\slashed E_T$    & \\  
  0.02 & $\ell$ 2$q$ $\slashed E_T$    & \\  
  0.02 & 2$\ell$ 2 $\tau$ $\slashed E_T$    & \\  
 \end{tabular}}
\scalebox{0.75}{\begin{tabular}{c|c | c}
BR & Channel & SS  \\
\hline
 0.01 &  2$\ell$   2$q$  $\slashed E_T$    & \\  
 0.01 & 3$\ell$   2$q$  $\slashed E_T$    &  {\Large \checkmark} \\  
 0.01 & $\ell$ $\tau$ $\slashed E_T$    & \\ 
  0.01 & 3$\ell$ $\tau$ $\slashed E_T$    & \\ 
  0.01 & $\slashed E_T$    & \\  
 0.01 & 3$\tau$ 2$q$ $\slashed E_T$    & \\ 
  0.01 & 2$\tau$ 2$q$ $\slashed E_T$    & \\ 
  0.01 & 2$\ell$ $\slashed E_T$    & \\ 
  0 & 4$\ell$ $\slashed E_T$    &  {\Large \checkmark} \\ 
    0 & $\ell$ 3$\tau$ $\slashed E_T$    & \\    
    0 &  2$\tau$ $\slashed E_T$    & \\        
    0 &  4$\tau$ $\slashed E_T$    & \\         
   & \\
\end{tabular}}
\end{center}
\caption{Decay rates from $N N$ production for $\bf{BP1_{IH}}$ (left) and $\bf{BP2_{IH}}$ (right). The rates are obtained by summing on all charges and flavor configurations. Here $\ell=e,\mu$.
The checkmarks correspond to channels that can produce a SS lepton signal.}\label{tab:BP12_IH} 
\end{table}

\subsection{Heavy neutrinos lifetime}\label{sec:RH_lifetimes}

A crucial quantity affecting the phenomenology and thus the search strategies for RH neutrinos is their lifetime $\tau_N =1/\Gamma_N$. We focus for simplicity on the case of (almost) degenerate RH neutrinos and we start by considering only the decay modes of Sec.~\ref{sec:decay} that are induced at the renormalizable level by the active-sterile mixing~\footnote{These decay modes turns out to be the dominant ones also in the presence of $d=6$ operators when the minimal flavor violation paradigm is imposed~\cite{Barducci:2020ncz}.}.  In particular, as we will show in Sec.~\ref{sec:prompt_N}, one can have RH neutrinos that decay promptly, displaced or are stable on detector length scales.

\subsubsection*{Prompt decay}

We consider a RH neutrino decay as prompt if it happens within $\sim 0.1\;$cm from the primary vertex. The production of a pair of $N$ gives rise to a six-body final state, including  signatures with high lepton multiplicity, see Tab.~\ref{tab:BP12_NH} and Tab.~\ref{tab:BP12_IH}. As we will see, in order to have promptly decaying RH neutrinos, one needs to have a large breaking of the naive see-saw scaling between the active-sterile mixing, the RH neutrino and the light neutrino masses. In the notation of Sec.~\ref{sec:th} this breaking is parametrized by a large value of the $\gamma$ parameter, see Eq.~\eqref{eq:increase}. Large mixing angles are however constrained by a variety of experimental searches, and too large values of $\gamma$ are thus ruled out.

\subsubsection*{Displaced decay}

A particle is considered to decay displaced if it decays away from the primary vertex but within the detector environment. The precise distance for defining a vertex to be displaced clearly depends on the specific detector geometry. Given that our study focuses
on future proposed $e^+e^-$ colliders, for which  detailed detector characteristics  have not yet been settled, we consider as displaced particles decaying between  $0.1\,$cm and $1\,$m from the primary vertex. Given the preliminary nature of our study we also consider the detector to have a spherical symmetry, instead  of a cylindrical one.

\subsubsection*{Decays outside the detector}

Also in this case the precise value of the decay length of the RH neutrinos in order for it to be considered detector stable depends on the specific geometry of the detector. We then consider as detector stable, RH neutrinos which decay more than $5\,$m away from the primary vertex.


\section{Future Higgs Factories}\label{sec:colliders}

\subsection{Collider prototypes}

Lepton colliders are  ideal machines for SM precision measurements due to the cleanliness of their environment, the precise knowledge of the initial-state particles configuration and, for some prototypes, the possibility of having polarized beams that can help to enhance the signal-to-background ratio. Despite their center-of-mass energy being typically smaller than the one of hadronic machines, they however offer excellent prospects in the direct search for NP. This is mainly due to the low SM backgrounds, whose rates are generically comparable to the searched signals, as opposed to what happens in hadronic machines.
Amongst the various proposals for a new generation of colliders after the HL-LHC era, leptons colliders then stand out as one of the more concrete possibility.

Various $e^+ e^-$ prototypes, presently at different stages of their design, have been proposed. These include circular ones, as the Future Circular Collider (FCC-ee)~\cite{Gomez-Ceballos:2013zzn,Abada:2019zxq,Abada:2019lih,Blondel:2019yqr}  and the Circular Electron Positron Collider (CEPC)~\cite{CEPCStudyGroup:2018rmc,CEPCStudyGroup:2018ghi}, and linear ones, as the International Linear Collider (ILC)~\cite{Bambade:2019fyw,Behnke:2013xla,Baer:2013cma} and the Compact Linear Collider (CLIC)~\cite{deBlas:2018mhx,Roloff:2018dqu}. We report in Tab.~\ref{tab:colliders}  the center-of-mass energies and luminosities considered in this study for various benchmark configurations. Notice that for CLIC we focus on its 
low-energy stage  at $\sqrt s=380\;$GeV, to which we refer as CLIC-380 throughout the text.
We report both the parameters for Higgs physics runs as well as the ones for runs at $\sqrt s=m_Z$. For what concerns the Higgs-strahlung cross sections $\sigma(e^+ e^- \to Zh)$, in the cases of the ILC and CLIC-380, these are reported under the assumptions of a beam polarization fraction $({\cal P}_{e^-},{\cal P}_{e^+})$ of  $(-80\%,+30\%)$ and $(-80\%,0\%)$, respectively.

\begin{table}[t!]
\begin{center}
\scalebox{0.85}{
\begin{tabular}[t]{ c || c | c | c   }
        \multicolumn{4}{c}{{\bf{Higgs run}}}	        \vspace{0.2cm} \\
 Collider & $\sqrt s\;$ [GeV] & $\int {\cal L}\;$[ab$^{-1}$] & $\sigma_{Zh}$ [fb]  \\ 
 \hline
 \hline
  \multirow{1}{*}{FCC-ee} & 240 & 5 & 193 \\
\hline
    \multirow{1}{*}{ILC} 	& 250 & 2 (pol) & 297 \\
\hline
       CLIC-380& 380 &1 (pol) & 133  \\	
    \hline	  
        CEPC& 240 &  5.6&  193  \\		
\end{tabular}
\hspace{0.8cm}
\begin{tabular}[t]{ c   || c | c | c  }
        \multicolumn{4}{c}{{\bf{$Z$ pole run}}}	        \vspace{0.2cm} \\
 Collider & $\sqrt s\;$ [GeV] & $\int {\cal L}\;$[ab$^{-1}$] & $N_{Z}$ \\ 
 \hline
 \hline
  \multirow{1}{*}{FCC-ee} &  $m_Z$ & 150 & $6.5\times 10^{12}$ \\
\hline
        CEPC & $m_Z$ & 16 & $6.9 \times 10^{11}$ \\		  
\end{tabular}
}
\end{center}
\caption{Center-of-mass energies and total integrated luminosities for the various collider options considered in the analysis for Higgs runs (left) and $Z$ pole runs (right). For the Higgs runs we report the values of the Higgs-strahlung cross-section reported in~\cite{deBlas:2019rxi}, while for the $Z$ pole runs the number of expected $Z$ bosons produced with the corresponding integrated luminosity from~\cite{ALEPH:2005ab}.}
\label{tab:colliders}
\end{table}

\subsection{Simulation Framework}

In our  analysis, signal events have been simulated at parton level by {\tt MadGraph5\_aMC@NLO}~\cite{Alwall:2014hca}. 
 The events have then been analysed with the {\tt MadAnalysis5} package~\cite{Conte:2012fm,Conte:2014zja,Dumont:2014tja}.
We consider our sensitivity estimates to be preliminary, not including any 
irreducible or reducible backgrounds. 
Nevertheless, we expect them to be not too far 
from  a realistic  lepton-collider sensitivity projection. For instance,  
an Higgs invariant-mass selection cut 
on the final states featuring 
very high multiplicity Higgs decays, as we are considering here, should 
be sufficient for suppressing  the irreducible SM backgrounds in the relatively clean lepton-collisions
environment. Although a full simulations of the actual detector performance, when available, will certainly make the corresponding projections more robust, we are confident that, 
in a more realistic approach, the 
excellent accuracy of {\it particle-flow} reconstruction, as now under consideration for Higgs Factories detectors, 
complemented by advanced analysis techniques,  might only moderately  degrade  the present sensitivity estimates.

\section{The ${\cal O}_{NH}$ operator and the Higgs-strahlung channel}\label{sec:ONH}

We start our analysis by discussing the ${\cal O}_{NH}$ operator. It can be generated by the tree-level exchange of a scalar singlet or by a fermion doublet with hypercharge $\pm 1/2$~\cite{Aparici:2009fh}. It gives rise to a new interaction of the Higgs boson with a pair of RH neutrinos. If kinematically allowed, this interaction induces an extra decay channel for the Higgs boson, $h\to NN$. In the following we assume for simplicity degenerate RH neutrino masses $m_{N_1}=m_{N_2}=m_N$. For real couplings~\footnote{For imaginary couplings there is a different dependence on the RH neutrino velocity $\beta_N$ due to CP properties of the matrix element see, {\emph{e.g.}},~\cite{Graesser:2007yj}.} the partial width reads~\cite{Graesser:2007yj}
\be\label{eq:GammaONH}
\Gamma( h \to \bar N^{c}_i N_i)=\frac{1}{2 \pi}\frac{ v^2}{\Lambda^2} m_H \beta_{N}^3 (\alpha_{NH}^{ii})^2 \ ,
\ee
where
\be
\beta_{N} = \sqrt{1-\frac{4 m_{N}^2}{m_H^2}} \ .
\ee
This operator can be constrained by searches for additional  untagged 
Higgs decay modes or invisible Higgs decays~\cite{deBlas:2019rxi}. More importantly, it adds a new production mode for RH neutrinos through the Higgs-strahlung process
\be\label{eq:higgs_strah}
e^+ e^- \to Z h, h \to N N \ .
\ee

At lepton colliders the process of Eq.~\eqref{eq:higgs_strah} is the dominant production mode for a SM Higgs boson for center-of-mass energies $\sqrt{s}\lesssim 400\;$GeV. It is crucial that, by using the recoil mass technique, this process can be tagged  by reconstructing the $Z$ decay products, without any knowledge of the particles arising from the Higgs boson decay. This property makes this channel important for all the three regimes of the RH neutrinos lifetimes described in Sec.~\ref{sec:RH_lifetimes}. The Higgs-strahlung cross-sections for the various colliders are normalized as reported in Tab.~\ref{tab:colliders}.

\subsection{Prompt decay}\label{sec:prompt_N}

\begin{figure}[t!]
\begin{center}
\includegraphics[width=0.60\textwidth]{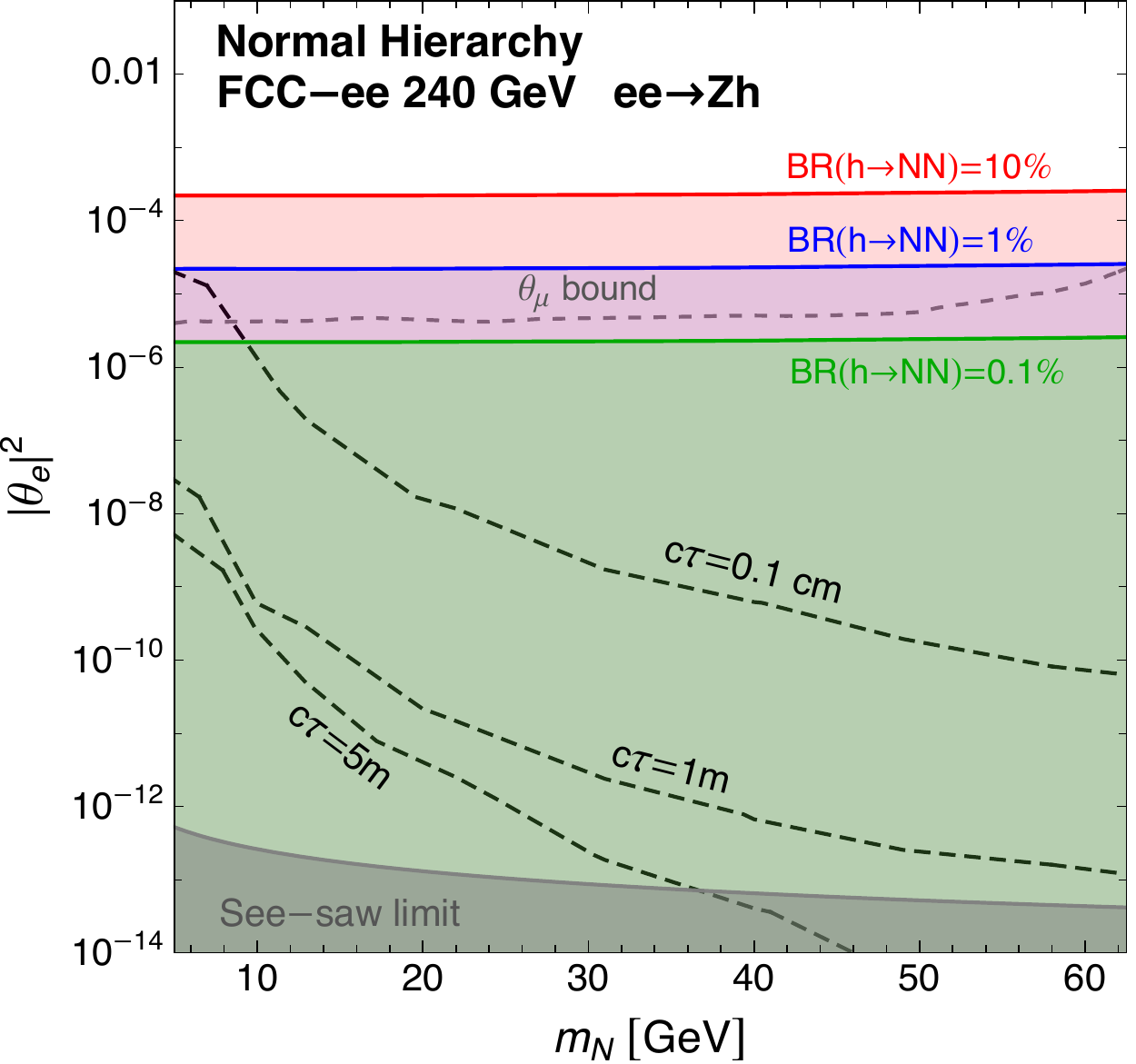}\hfill
\caption{\small 
Regions in the $m_N-|\theta_e|^2$ parameter space where the RH neutrino pair production from the Higgs-strahlung process of Eq.~\eqref{eq:higgs_strah} is larger than the production of a single RH neutrinos via mixing, for ${\rm BR}(h\to NN)=10\%$ (red), 1\% (blue) and 0.1\% (green) for the FCC-ee case. The NH case is assumed. The gray dashed line represents the limit on the mixing angle arising from existing experimental searches,  while the black dashed lines represent isocontour of proper decay length $c\tau$. In the gray shaded region  the lightness of the neutrino masses cannot be explained by the see-saw mechanism.}
\label{fig:mixing_vs_ONH}
\end{center}
\end{figure}

As anticipated in Sec.~\ref{sec:mixing-formalism}, in order to have the RH neutrinos to decay promptly, one needs a large breaking of the naive see-saw scaling, parametrized by a large value of the parameter $\gamma$ entering the Casas-Ibarra parametrization of Eq.~\eqref{eq:Rmatrix} and enhancing the active-sterile mixing angle.  This mixing is however constrained by a variety of experimental searches and too large values of $\gamma$ are ruled out. Using the bounds on $\theta_\alpha = \sum_{i=1,2} |\theta_{\alpha i}|^2$ reported in~\cite{Liventsev:2013zz,Aaij:2016xmb, Abreu:1996pa}, we show in Fig.~\ref{fig:mixing_vs_ONH} as a gray dashed line the exclusion contour in the $m_N-|\theta^2_e|$ plane for the NH case. Similar results are obtained for IH. We show only the bound arising from $\theta_\mu$, which turns out to be the most stringent one. Notice that $|\theta_e|^2$ inherits the bound from $\theta_\mu$ through its dependence on $\gamma$ and $m_N$. In the same plot we also show as black dashed lines the isocontours of proper decay length $c\tau$ of the RH neutrino $N_1$, in order to identify the regions where the decay is prompt, displaced or outside of the detector~\footnote{Strictly speaking, the important quantity is the laboratory frame decay length, which is larger than the proper decay length due to Lorentz time dilation. We will accurately compute this quantity in Sec.~\ref{sec:ONH-disp} when dealing with displaced vertices while in this section we assume to be in a region of the $m_N-|\theta^2_e|$ parameter space where the Lorentz factor does not modify the behavior between prompt, displaced or stable. This is the case for $c\tau$ values away from the boundary regions indicated in the plot for not too light $N$.}. We neglect the dependence of the lifetime on $\beta$, $\delta$ and $\phi_1$ which is mild for $\gamma \gtrsim 1$. We also show the colored regions in which the RH neutrino pair production from the Higgs-strahlung process of Eq.~\eqref{eq:higgs_strah} is larger than the production of a single RH neutrinos via mixing. For concreteness we consider the case of the FCC-ee collider with $\sqrt{s}=240\;$GeV, and compare the Higgs-strahlung cross section with the one of production via mixing. Since the result depends on the branching ratio of the Higgs boson into RH states, we show our results for ${\rm BR}(h\to NN)=10\%$ (red), 1\% (blue) and 0.1\% (green), using~\cite{Antusch:2016vyf} for the normalization of the mixing cross-section. Finally, in the gray area at the bottom of the plot the lightness of the neutrino masses cannot be explained by the see-saw mechanism. 

Altogether, the figure makes clear that there are large regions in parameter space in which  Higgs-strahlung production may dominate over mixing production. Moreover, depending on $m_N$, in these regions the RH neutrinos can have prompt, displaced or outside the detector decays depending on the active-sterile mixing.

\subsubsection{Projected sensitivities: $e$ and $\mu$ mixing}\label{sec:res_2l4q_prompt}

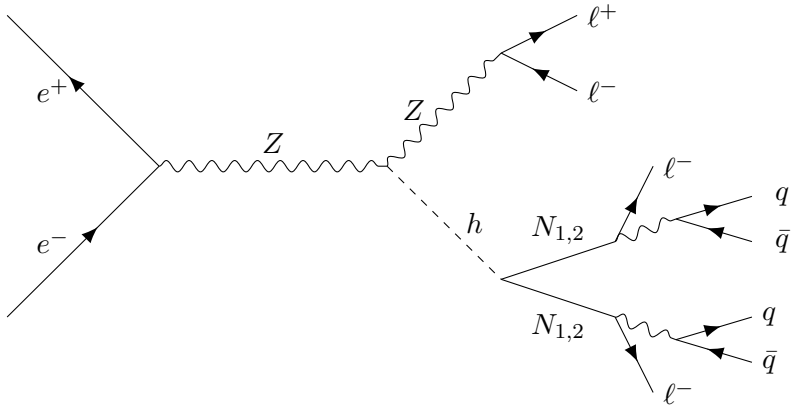
\begin{figure}
\begin{center}
\adjustbox{valign=m}{
 \begin{tikzpicture}
\draw[f] (-2,-2) -- (0,0) node[midway, xshift=-4mm]{$e^-$};
\draw[f] (0,0) -- (-2,2) node[midway, xshift=-4mm]{$e^+$};
\draw[v] (0,0) -- (3,0) node[midway, yshift=3mm]{$Z$};
\draw[v] (3,0) -- (4.5,1.5) node[midway,xshift=-4mm]{$Z$};
\draw[f] (5.5, 1) node[right]{$\ell^-$} -- (4.5,1.5) ;
\draw[f] (4.5,1.5) -- (5.5,2) node[right]{$\ell^+$};
\draw[dashed] (3,0) -- (4.5,-1.5) node[midway, xshift =4mm]{$h$};
\draw (4.5,-1.5) -- (6, -1) node[midway,yshift=4mm]{$N_{1,2}$};
\draw (4.5,-1.5) -- (6,-2) node[midway,yshift=-4mm]{$N_{1,2}$};
\draw[f] (6,-1) -- (6.5, 0) node[right]{$\ell^-$};
\draw[v] (6, -1) -- (6.8, -0.7);
\draw[f] (6.8, -0.7) -- (7.8, -0.4) node[xshift=4mm]{$q$};
\draw[f] (7.8, -1) node[xshift=4mm]{$\bar q$} -- (6.8, -0.7) ;
\draw[f] (6,-2) -- (6.5, -3) node[right]{$\ell^-$};
\draw[v] (6,-2) -- (6.8, -2.3);
\draw[f] (6.8, -2.3) -- (7.8, -2) node[right]{$q$};
\draw[f]  (7.8, -2.6) node[right]{$\bar q$} -- (6.8, -2.3);
\end{tikzpicture}} 
\end{center}
\caption{Feynman diagram for $2\ell 4q$ production in the SS leptons final state through Higgs-strahlung production. }\label{fig:feynman_ONH}
\end{figure}

We start by considering the benchmark points which minimize the mixing with the third generation leptons, namely ${\bf BP1_{NH}}$ and ${\bf BP1_{IH}}$ of Eq.~\eqref{eq:BP_NH} and Eq.~\eqref{eq:BP_IH}. For these two cases we focus for simplicity on the final state with the highest rate, which according to Tab.~\ref{tab:BP12_NH} and Tab.~\ref{tab:BP12_IH} is the $2\ell 4q$ one. Furthermore, we require this final state to contain a pair of SS leptons, thus halving the rates reported in the tables, and focus on the case in which the $Z$ boson decays leptonically. In computing our limits we sum on both $e$ and $\mu$ flavor combinations for the $Z$ and the $N$ pair decay modes, and on both the RH neutrinos $N_1$ and $N_2$. All together the process we analyze is
\be\label{eq:2l4q_SS}
e^+ e^- \to Z h \to (\ell_\alpha^+ \ell_\alpha^- )( \ell_\beta^+ \ell_\gamma^+ 4q) + h.c.
\ee
where the first bracket indicates the $Z$ boson decay products, while the second bracket the Higgs boson ones and $\alpha, \beta,\gamma=1,2$ are flavor indices. This is shown in Fig.~\ref{fig:feynman_ONH}.
 For our analysis we require the leptons to have $p_T>2.5\;$GeV and $|\eta|<2.44$, while jets should satisfy $p_T>5\;$GeV and $|\eta| <2.4$. Leptons are required to be separated by $\Delta R>0.15$ among themselves and with respect to the selected jets. In order to tag the Higgs-strahlung topology, we require two same-flavor opposite-sign leptons with an invariant mass $|m_{\ell^+\ell^-}-m_Z|<10\;$GeV. If more than one pair that satisfies this condition is present, we choose the pair with an invariant mass closer to the $Z$ mass.  Furthermore this pair is also required to have a recoil mass $m_{{\rm rec}}$ within 10 GeV of the true Higgs mass $m_H$, where
\be
m^2_{{\rm rec}}  =s - 2\sqrt s E_{\ell^+ \ell^-}+m_{\ell^+ \ell^-}^2 \ ,
\ee 
and $E_{\ell^+ \ell^-}$ is the energy of the leptons pair.
We then ask for exactly  two other SS leptons in the event, while no requirements on the number of jets and missing energy is imposed.

\begin{figure}[t!]
\begin{center}
\includegraphics[width=0.50\textwidth]{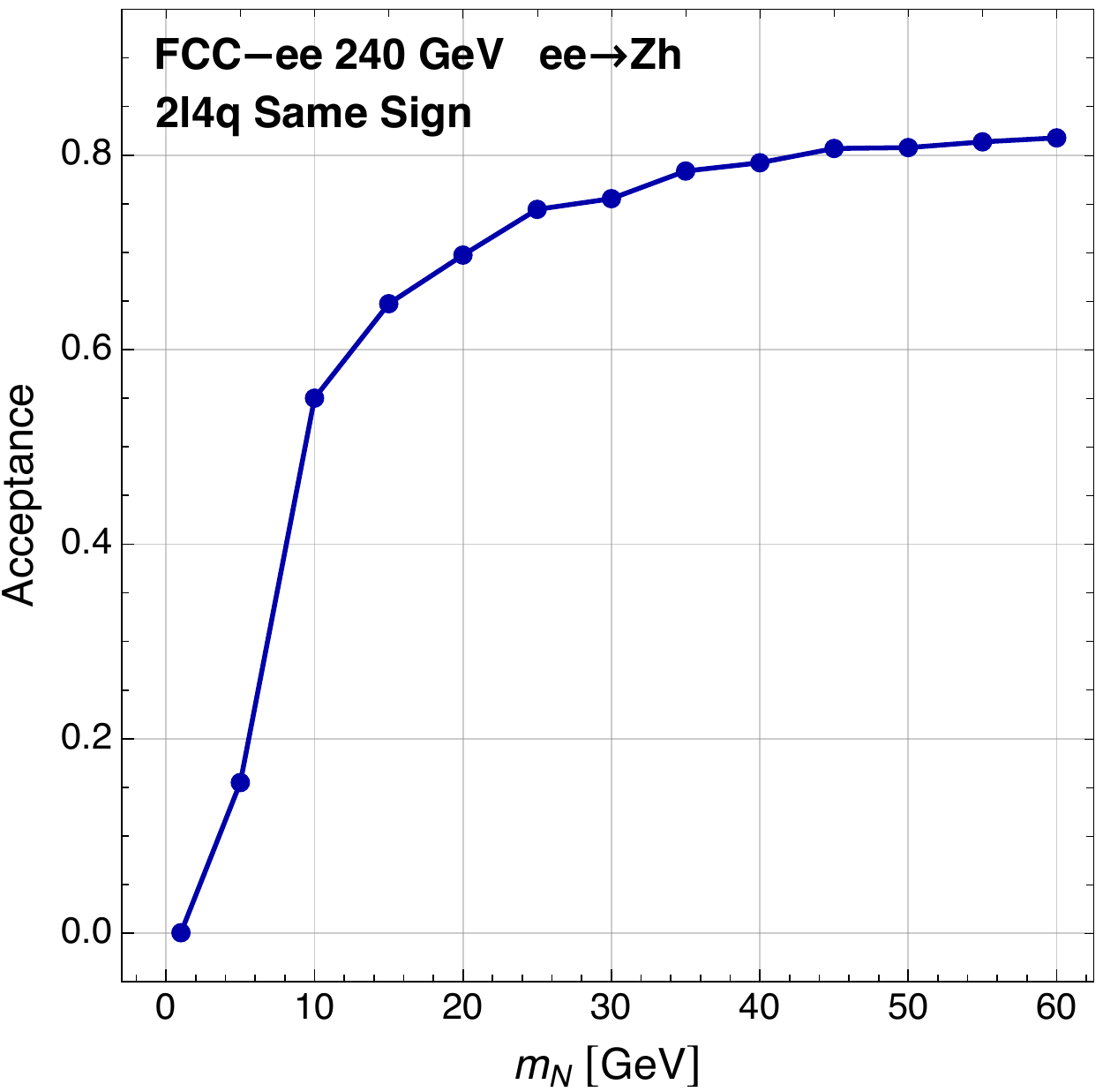}\hfill
\caption{\small Parton level acceptances for the $2\ell 4q$ final state in the  Higgs-strahlung topology with a leptonically decaying $Z$ boson and with the selection cuts described in the main text.}
\label{fig:acc_2l4q_SS_parton}
\end{center}
\end{figure}

Three final state flavor configurations from the  $N N$ decay contribute to the selected final state: $e^+e^+$, $e^+\mu^+$ and $\mu^+\mu^+$ plus the charge conjugated processes. The parton level acceptances as a function of the RH neutrino mass~\footnote{The acceptances are computed for the fully electron final state. Similar acceptances are obtained for the other flavor combinations.} for these  final states are shown in Fig.~\ref{fig:acc_2l4q_SS_parton}. As it can be seen the acceptances increase with the increase of the RH neutrino mass. This is to be expected, since for light RH neutrinos their decay products turn out to be more collimated, thus making it harder to pass the $\Delta R>0.15$ isolation criteria.

Regarding the irreducible SM backgrounds to the process of Eq.~\eqref{eq:2l4q_SS}, we expect it to be 
negligible.  
In principle,
 since we are not requiring any  jet in the final state, 
 any SM process matching a signature $(Z\to \ell_\alpha^+ \ell_\alpha^-)(h \to \ell_\beta^+ \ell_\gamma^+ + \dots)$ could mimic the signal. Lepton-number conservation in the SM however implies that at least two extra neutrinos should be present in the final state. Electric-charge conservation  moreover implies that  there should also be at least two further quark pairs, arising {\emph{e.g.}} from two off-shell $W$'s.
A typical SM irreducible background to the process under consideration is therefore
\be
e^+ e^- \to (Z\to \ell_\alpha^+ \ell_\alpha^-)(h \to \ell_\beta^+ \nu_\beta \ell_\gamma^+ \nu_\gamma \bar u d \bar u d)
\ee
where the final state arising from the Higgs decay involves multiple off-shell gauge bosons, such a $h\to 4W^\ast$ decay. We then expect it to be totally negligible. Due to the very large multiplicity of the chosen signal, we also expect  other channels that match the same final state without occurring via  $Z$ and Higgs resonances to be negligible.
As for other reducible backgrounds arising from the limited efficiencies in the detector performances, such as {\emph{e.g.}} mis-identification of the
final particles, we are confident that they can be efficiently removed thanks to the strong kinematical characterization of the  $(Z\to \ell_\alpha^+ \ell_\alpha^-)(h \to \ell_\beta^+ \ell_\gamma^+ + \dots)$ process.
 We thus assume  the  process in Eq.~\eqref{eq:2l4q_SS} to be background free. 
  According to Poisson statistics,  in case no signal event is
  observed,
  one can then  set a 95\% confidence-level (CL) exclusion limit corresponding to  a maximally allowed number $N_s=3$ of new physics  events~\cite{Zyla:2020zbs}.

\begin{figure}[t!]
\begin{center}
\includegraphics[width=0.49\textwidth]{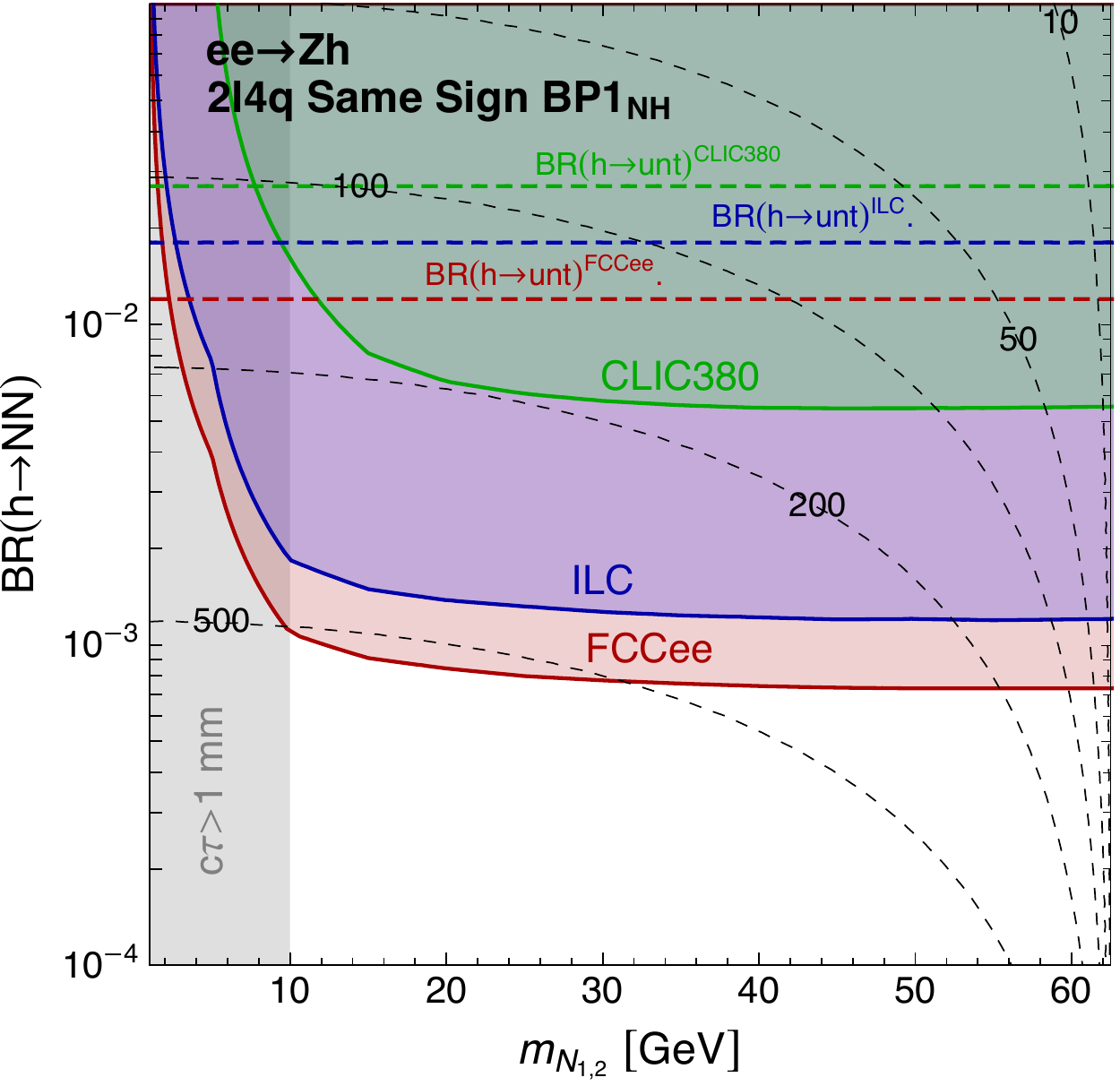}\hfill
\includegraphics[width=0.49\textwidth]{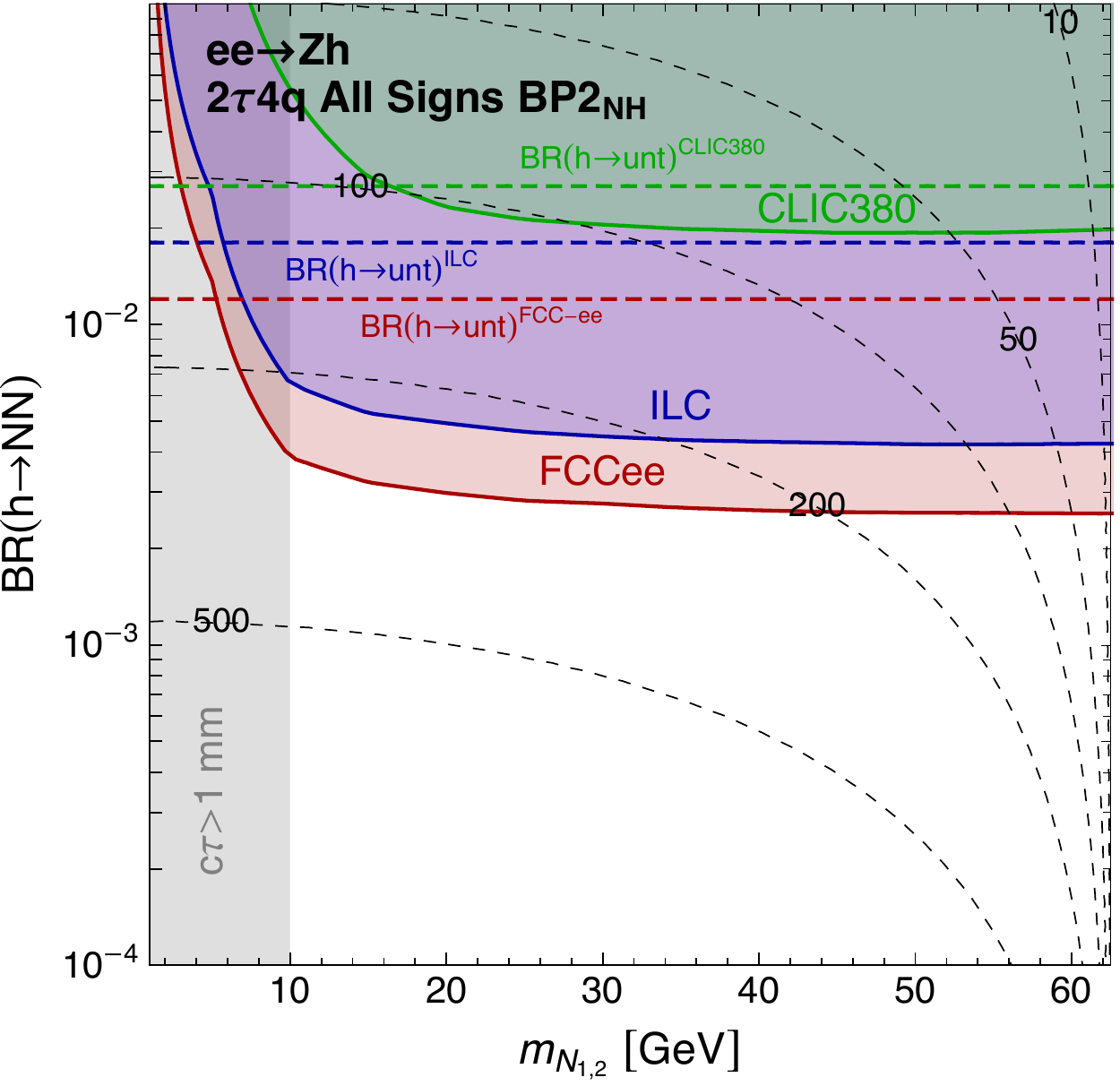}
\caption{\small 95\% CL exclusion for ${\bf BP1_{NH}}$ (left) and ${\bf BP2_{NH}}$ (right) from prompt searches in the Higgs-strahlung channel. The shaded areas represent the exclusion contours for the case of FCC-ee (red), ILC (blue) and CLIC-380 (green) adopting the strategy of Sec.~\ref{sec:res_2tau4q_prompt} and Sec.~\ref{sec:res_2tau4q_prompt} for the case of the $2\ell 4q\;({\rm SS})$ and $2\tau 4q$ final states respectively. Also shown as colored dashed lines the limits arising from untagged Higgs decay measurements and as gray dashed lines the isocontours of NP scale $\Lambda$ in TeV. In the gray shaded area $c\tau >1\;$mm and the RH neutrino cannot decay promptly without being excluded by experimental searches.}
\label{fig:2l4q_SS_opt}
\end{center}
\end{figure}

We show our results for  ${\bf{BP1}_{NH}}$ in the left panel of Fig.~\ref{fig:2l4q_SS_opt}. Similar results are obtained for the ${\bf{BP1}_{IH}}$ benchmark. The colored regions represent the 95\% CL sensitivity of the analysis on the exotic Higgs BR described above in the $m_N - {\rm BR}(h \to NN)$ plane for the various collider options. This reach has to be compared with the limits coming from untagged Higgs decay~\cite{deBlas:2019rxi}, represented by horizontal dashed colored lines. For simplicity we do not show the limit that might be obtained at CEPC, which are comparable to the ones from FCC-ee, due to their similar integrated luminosity, see Tab.~\ref{tab:colliders}. We also show in dashed black the isocontours of the scale $\Lambda$ expressed in TeV, fixing $\alpha_{NH}^{ii} = 1$. Finally, we highlight in gray the region where $c\tau>1\;$mm in which the RH neutrino cannot decay promptly  without being excluded by experimental searches, see Fig.~\ref{fig:mixing_vs_ONH}.

All together we see that future $e^+e^-$ colliders will be able to set a bound on the exotic Higgs BR that goes from $\sim 5 \times 10^{-3}$ for the case of CLIC-380 down to $\sim 7 \times 10^{-4}$ for the case of FCC-ee/CEPC, and that these bounds are significantly stronger than the corresponding ones arising from untagged Higgs decay measurements. In terms of NP scale $\Lambda$ these limits translate in a bound which, in the most favorable case, is $\Lambda \gtrsim 500\;$TeV.

\subsubsection{Projected sensitivities: $\tau$ mixing}\label{sec:res_2tau4q_prompt}

We now consider the case where the active-sterile mixing with the third generation of leptons is maximized, as is the case for the benchmarks ${\bf BP2_{NH}}$ and ${\bf BP2_{IH}}$ in Eq.~\eqref{eq:BP_NH} and Eq.~\eqref{eq:BP_IH}. Again we choose for simplicity the channel with the highest BR for both benchmark points, {\emph{i.e.}} the $2\tau4q$ channel. In this case however the $\tau$ leptons will promptly decay into a $\nu_\tau$ and an off-shell $W$ boson. For our analysis we adopt the following strategy. We first consider the $\tau$ leptons as stable particles, and apply the same parton-level selection efficiencies computed in the $e$ and $\mu$ case of Sec.~\ref{sec:res_2l4q_prompt}.  Then we focus on the hadronic decay modes $\tau^- \to \pi^- \pi^0 \nu_\tau$, $\tau^- \to \pi^- \nu_\tau$ and $\tau^- \to \pi^- \pi^0 \pi^0 \nu_\tau$ (which account, respectively, for approximately $25.5\%$, $10.8\%$ and $9.3\%$ of the total $\tau$  branching ratio) and apply, following~\cite{Tran:2015nxa}, a flat 90\% reconstruction efficiency for each $\tau$ lepton. We consider  the inclusive $2\tau 4q$ final state with all charge combinations, $\tau^\pm \tau^\pm$ and $\tau^\pm \tau^\mp$, and, again, neglect the SM background, although in this case
we expect a larger contamination with respect to the $2\ell 4q\;{\rm SS}$ 
case~\footnote{For instance we estimated the cross section for the irreducible background 
 $e^+ e^-  \to ( Z \to \mu^+ \mu^-) (h \to \tau^+ \tau^- u \bar d \bar u d)$ 
  to be about $5.4 \times 10^{-6}\;$pb, corresponding to about 27 events at FCC-ee. We expect that these can be efficiently reduced by exploiting the kinematical features of the $h\to NN$ decay.
   }. Under these assumptions we obtain the results shown in the right panel of Fig.~\ref{fig:2l4q_SS_opt} for ${\bf BP2_{NH}}$. Similar, albeit slightly weaker, results are obtained for the ${\bf BP2_{IH}}$ case. We see that also in the case in which $N$ mixes dominantly with $\nu_\tau$, both FCC-ee and ILC can probe values of BR($h\to NN$) that go well beyond the limits arising from untagged Higgs decay, while CLIC-380 can marginally surpass this reach.

\begin{figure}[tb]
\begin{center}
\includegraphics[width=0.48\textwidth]{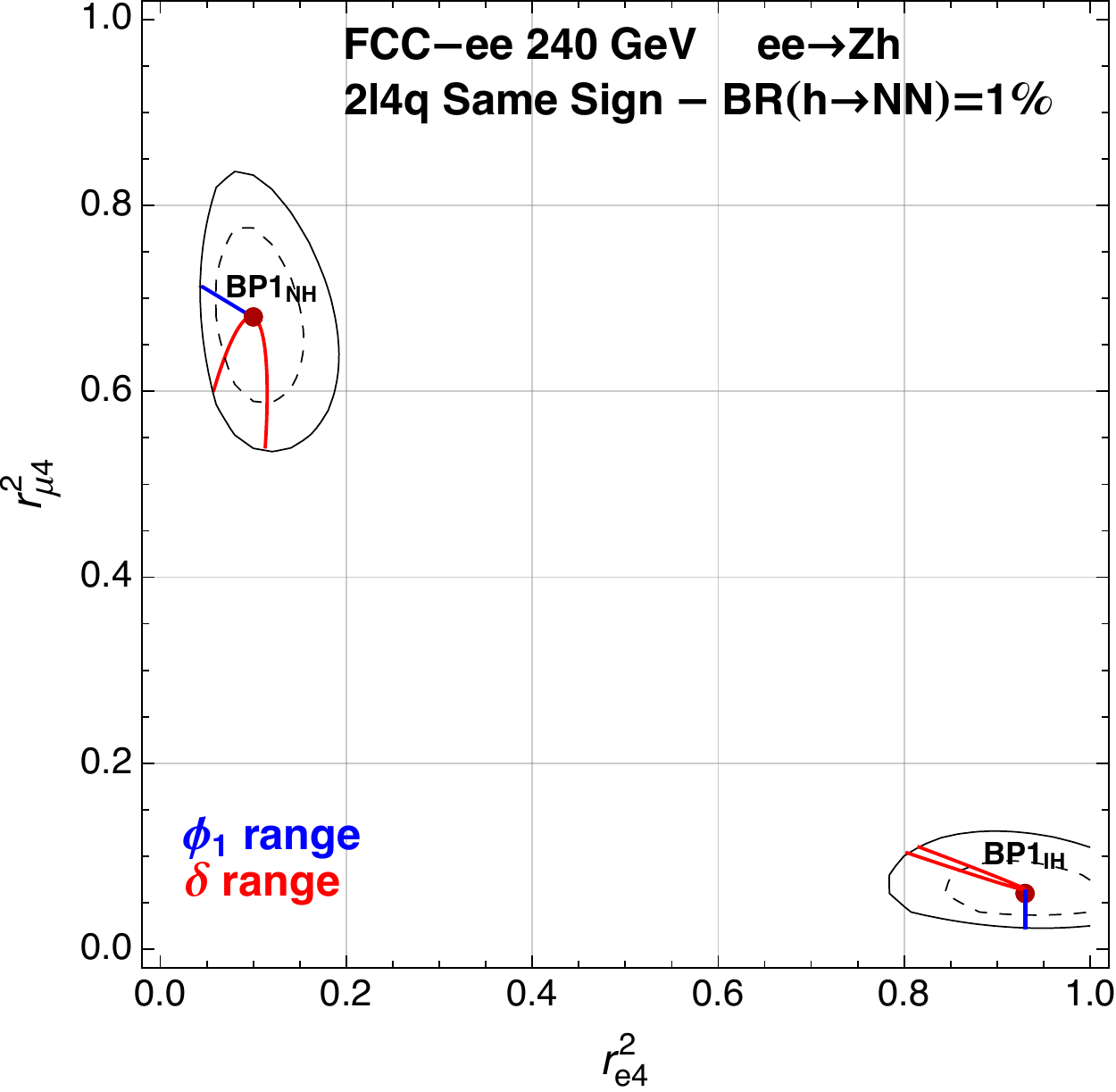}
\includegraphics[width=0.48\textwidth]{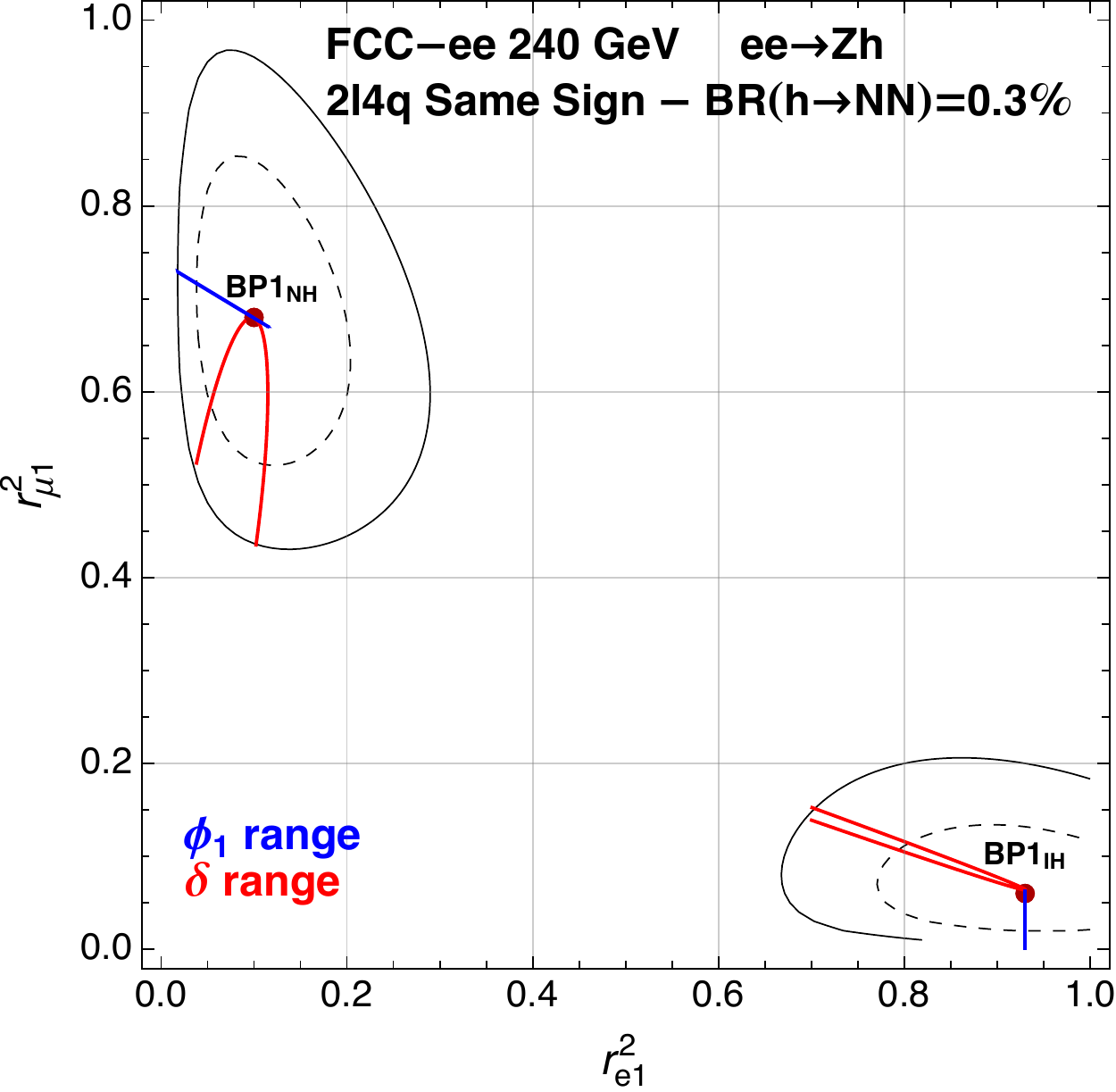}
\caption{ 68\% (dashed) and 95\% (solid) confidence intervals for determining the flavor structure of a putative BSM signal observed associated with ${\bf BP1_{NH}}$ and ${\bf BP1_{IH}}$ in the $2\ell 4q$ final state with a pair of SS leptons. We fix ${\rm BR}(h\to NN)=1\%$ (left), $0.3\%$ (right) and $m_N=30\;$GeV.}
\label{fig:discrimination}
\end{center}
\end{figure}
\subsubsection{Determination of the flavor structure}\label{sec:res_prompt_determination}

Should a RH neutrino signal be detected, a crucial question to be asked is with which accuracy it will be possible to determine the flavor structure of the underlying BSM theory. This is directly related to the number of signal events which in turn depends on the physics of the underlying theory, namely ${\rm BR}(h \to NN)$, and the detector performances in reconstructing the signal. To quantify this we adopt the following strategy. We choose the $2\ell 4q$ final state with a pair of SS leptons, a representative RH neutrino mass of 30\;GeV and the benchmark points  ${\bf BP1_{NH}}$ and  ${\bf BP1_{IH}}$.  We then build a Poisson distribution $p(n_{{\rm obs}}|n_{{\rm th}})=\frac{1}{n_{{\rm obs}}!} e^{-n_{{\rm th}}} n_{{\rm th}}^{n_{{\rm obs}}}$ where $n_{{\rm obs}}$ correspond to the expected number of  observed events in the case of the presence of a BSM signal, and $n_{{\rm th}}$ is the theoretical prediction for the number of events in the chosen final state, which is a function of $r^2_{e1}$ and $r^2_{\mu1}$. For fixed values of BR($h\to NN$), which then fixes the final event yield, we thus compute the 68\% and 95\% CL confidence interval around the chosen benchmark points. The results are illustrated in Fig.~\ref{fig:discrimination} for the case of ${\rm BR}(h\to NN)=1\%$ (left) and 0.3\% (right) for the FCC-ee collider option. We see that with a 1\% Higgs exotic BR, which correspond to $\sim 37$ total signal events in the final state, the normalized squared mixings $r_{e1}$ and $r_{\mu1}$ can be determined with an absolute error of $\sim 0.1$, which rapidly degrades down to $\sim 0.3$ with a 0.3\% Higgs BR into a pair of RH neutrinos, for which one has $\sim 11$ signal events. It is interesting to interpret these results in terms of the phases appearing in the PMNS matrix, $\delta$ and $\phi_1$. The benchmark points we choose correspond to fixed values for both phases: $(\delta , \phi_1) = (0.76, 4.59)$ for the ${\bf BP1_{NH}}$ benchmark, and $(\delta , \phi_1) = (3.25, 1.60)$ for the ${\bf BP1_{IH}}$ benchmark. By keeping $\delta$ fixed and allowing $\phi_1$ to vary we obtain the blue lines in Fig.~\ref{fig:discrimination}, while by keeping $\phi_1$ fixed and allowing $\delta$ to vary we obtain the red lines. The interval that $\delta$ and $\phi_1$ can span inside the 95\% CL confidence intervals around the benchmark points values are reported in Tab.~\ref{tab:interval_phases}.

\begin{table}[tb]
\begin{center}
\begin{tabular}{c||c|c}
& ${\rm BR}(h\to NN)=1\%$ & ${\rm BR}(h\to NN)=0.3\%$ \\
\hline\hline
\multirow{2}{*}{${\bf BP1_{NH}}$} & $3.69 \leq \phi_1 \leq 5.57 $ & $0.037 \leq \phi_1 \leq 5.95$ \\
						   & $ 0.78 \leq \delta \leq 1.85 \; \cup \; 4.47 \leq \delta \leq 5.55$ & $0 \leq \delta \leq 2.53 \; \cup \; 3.80 \leq \delta \leq 2\pi$ \\
\hline
\multirow{2}{*}{${\bf BP1_{IH}}$} & $0.80 \leq \phi_1 \leq 2.31 $ & $0.51 \leq \phi_1 \leq 2.31$ \\
						   & $ 1.33 \leq \delta \leq 5.09$  & $0 \leq \delta \leq 2\pi $  \\
\end{tabular}
\end{center}
\caption{Range of parameters that can be probed in case of detection of a RH neutrino decay for the representative $m_N$ mass and ${\rm BR}(h\to NN)$ considered in Fig.~\ref{fig:discrimination}. The range for $\phi_1$ corresponds to the red lines in the figure, while the range for $\delta$ to the blue ones. In both case the other phase is kept fixed to the benchmark point value.}
\label{tab:interval_phases}
\end{table}

\subsection{Displaced decay}\label{sec:ONH-disp}

We now study the sensitivity for RH neutrinos decaying with a displacement which, as discussed in Sec.~\ref{sec:RH_lifetimes}, we take to be between 1\,cm and 100\,cm from the primary vertex, see Fig.~\ref{fig:mixing_vs_ONH}.
We consider decays into first and second generation leptons and we focus again on the $2\ell 4q$ final state, which is the one that maximizes the decay rate of the $NN$ pair. The final state therefore consists of two prompt same flavor opposite sign leptons from the $Z$ boson decay and the $2\ell 4q$ system. The final event yield is given by
\be\label{eq:displaced_yield}
N_{s}=\sigma_{Zh}\times {\rm BR}(Z \to \ell^+ \ell^-) \times {\rm BR}(h\to NN) \times {\rm BR}(NN \to 2\ell 4q) \times \epsilon_{Zh}
\times \epsilon_{P_{\Delta L}}^2 \times \epsilon^2_{\rm disp.} \times {\cal L} \ ,
\ee
where $\epsilon_{Zh}$ is the acceptance for reconstructing the $Z$ boson and the Higgs recoil mass from the two same flavor opposite sign prompt leptons. The parameter $\epsilon^2_{P_{\Delta L}}$ represents instead the acceptance for having both neutrinos decaying within a certain displacement from the primary vertex. This probability can be computed from the exponential decay law, taking into account the time dilation factor obtained by boosting the events from the RH neutrino rest frame to the laboratory frame. In practice we have computed, for each event and for each RH neutrino, 
 the Lorentz $\beta$ and $\gamma$ factors
\be
\beta = \sqrt{1-\frac{1}{\gamma^2}} \ , \qquad 
\gamma = \sqrt{1+\frac{|\vec{p}|^2}{M_N^2}} \ ,
\ee 
and assigned a probability for having the RH neutrino decaying at a distance $\Delta x =  x_f - x_i$
\begin{align}
{\cal P}(x_i,x_f) & = e^{-\frac{x_i}{\beta\gamma c \tau}} - e^{-\frac{x_f}{\beta\gamma c \tau}} \ .
\end{align}
We have then accepted events where both RH neutrino decays happened between 1\,cm and 100\,cm from the primary vertex.  Finally, with $\epsilon_{\rm disp.}$ we parametrize the acceptance for reconstructing the displaced decay, including
various detector inefficiencies, which will depend on the actual detector design 
and performances, and which therefore we assume as a free extra parameter
in the analysis. The irreducible SM background is expected to be negligible on the considered decay lenghts. 

\begin{figure}[t!]
\begin{center}
\includegraphics[width=0.48\textwidth]{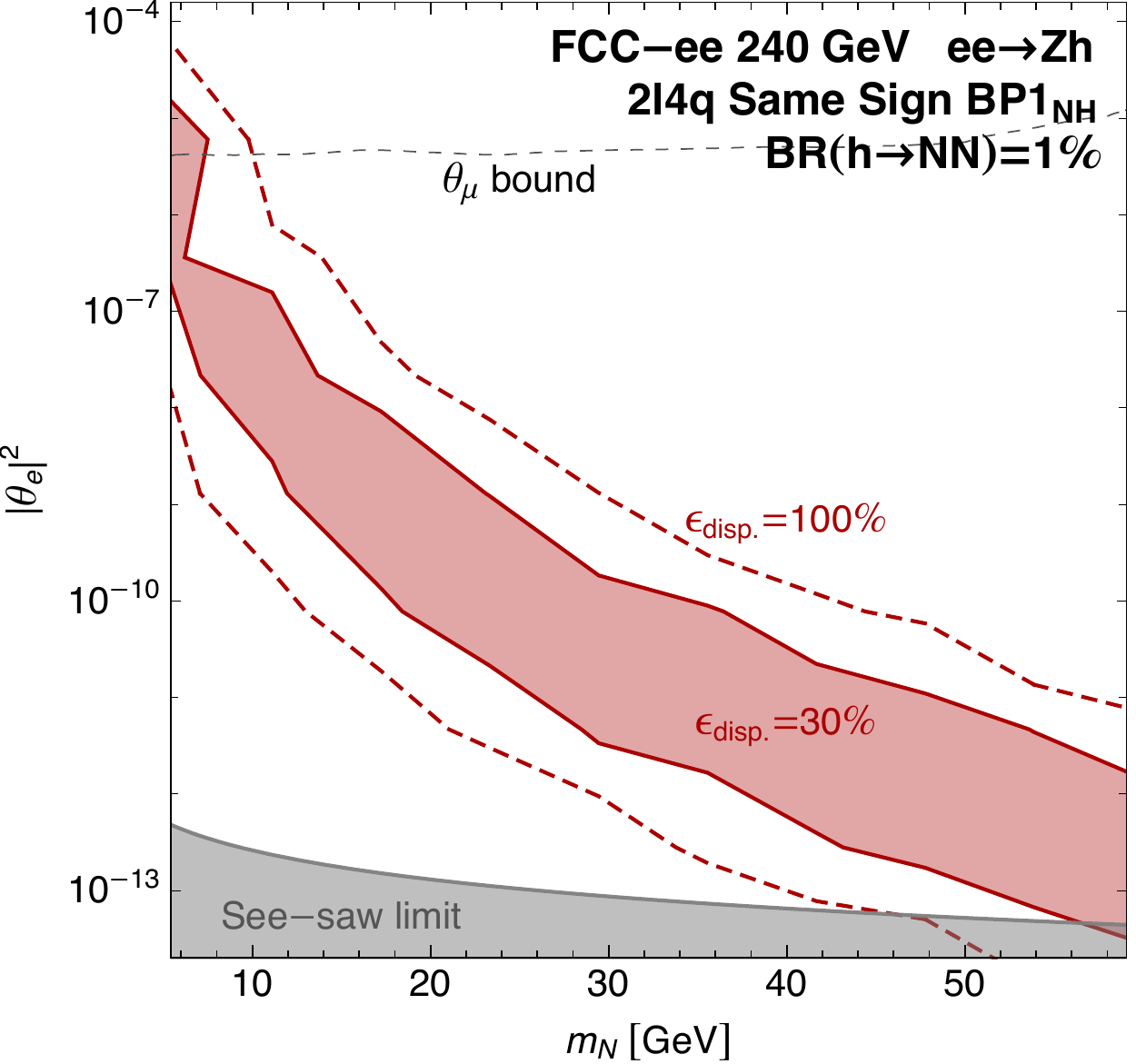}\hfill
\includegraphics[width=0.48\textwidth]{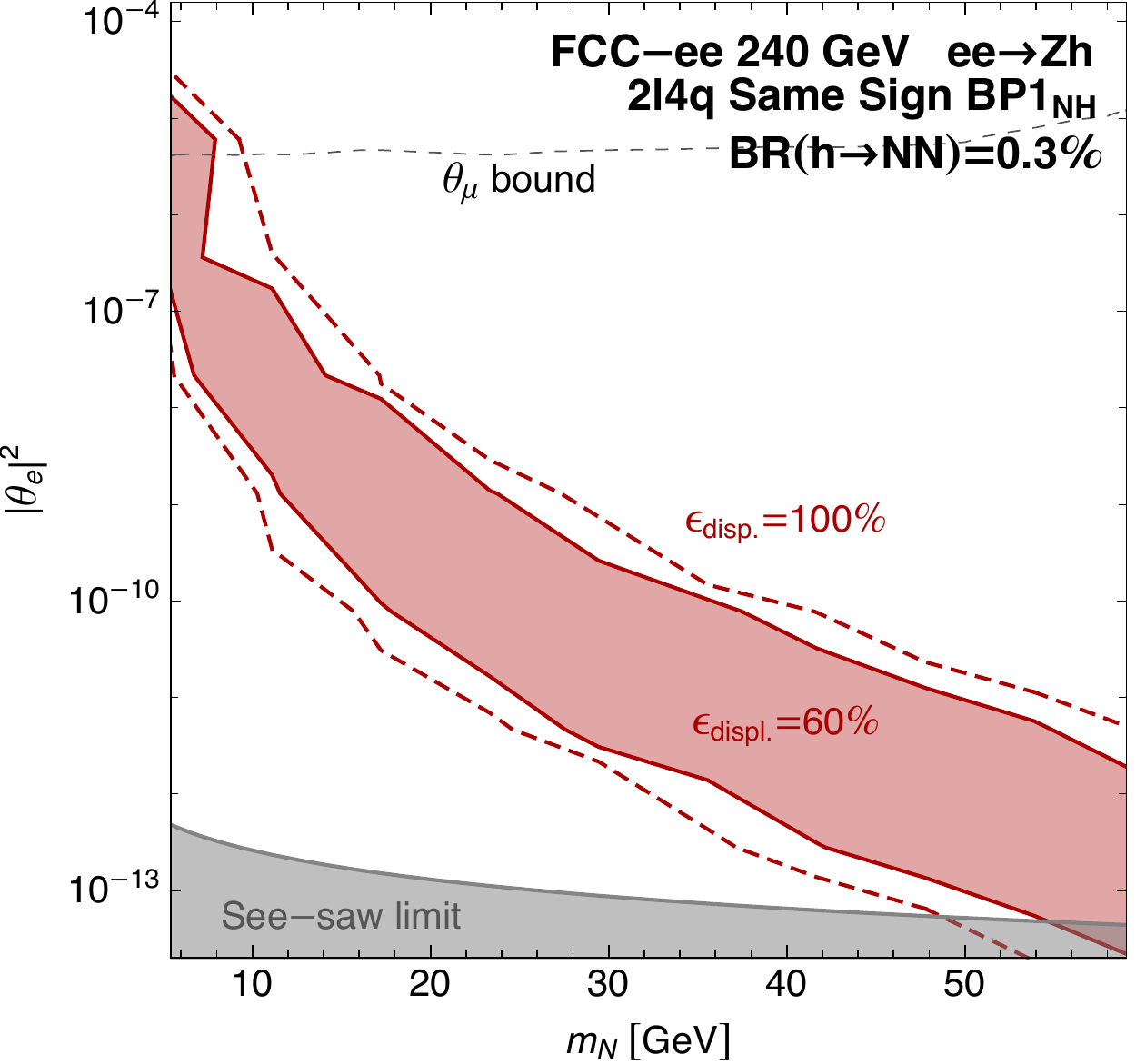}\hfill
\caption{
95\% CL exclusion for ${\bf BP1_{NH}}$ in the $m_N-|\theta_e|^2$ plane from displaced decay searches in the Higgs-strahlung channel
assuming ${\rm BR}(h\to NN)=1\%$ (left) and 0.3\% (right) for the case of FCC-ee. The shaded red area and the dashed red lines represent the sensitivities for different assumptions on the efficiencies for reconstructing the displaced vertices $\epsilon_{\rm disp.}$. The gray dashed line represents the limit on the mixing angle arising from existing experimental searches. In the gray shaded region  the lightness of the neutrino masses cannot be explained by the see-saw mechanism.
}
\label{fig:displaced_theta2}
\end{center}
\end{figure}

We  show in Fig.~\ref{fig:displaced_theta2} the FCC-ee reach for the ${\bf BP1_{IH}}$ benchmark in the $m_N-|\theta_e|^2$ plane. In the left panel we fix ${\rm BR}(h\to NN)=1\%$, while in the right panel we fix ${\rm BR}(h\to NN)=0.3\%$. We show our results for two different efficiencies $\epsilon_{\rm disp.}$ as reported in the plots. In both figures the gray shaded area represents the see-saw limit, while the values of the mixing angles above the gray dashed line are excluded by experimental searches, see the discussion in Sec.~\ref{sec:prompt_N}. Our results show that the search for RH neutrinos arising from Higgs decay and decaying displaced can offer a great handle in testing the active-sterile mixing angle. This is clearly due to the fact that, unlike in the case of production via mixing, the Higgs-strahlung cross section does not depend on the active-sterile mixing angle, which only enters in the lifetime determination, but only on the additional Higgs decay rate into an $NN$ pair. As a consequence, searches at FCC-ee could test values of $|\theta_e|^2$ down to the see-saw limit. We do not show for simplicity the limits arising from ILC and CEPC, which are comparable to the ones shown, and the one from CLIC-380, which turns out to be slightly weaker.

\begin{figure}[t!]
\begin{center}
\includegraphics[width=0.48\textwidth]{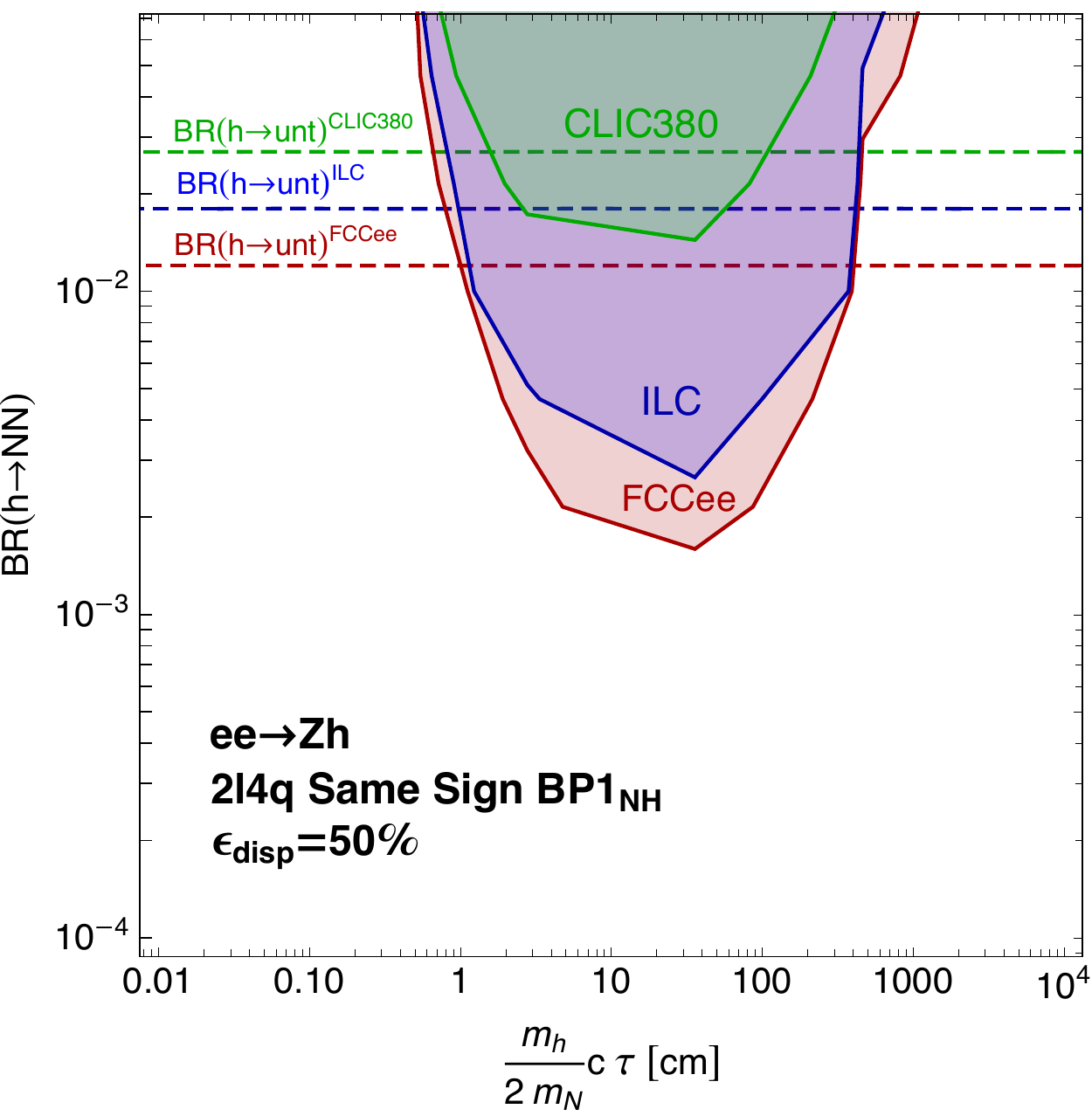}\hfill
\includegraphics[width=0.48\textwidth]{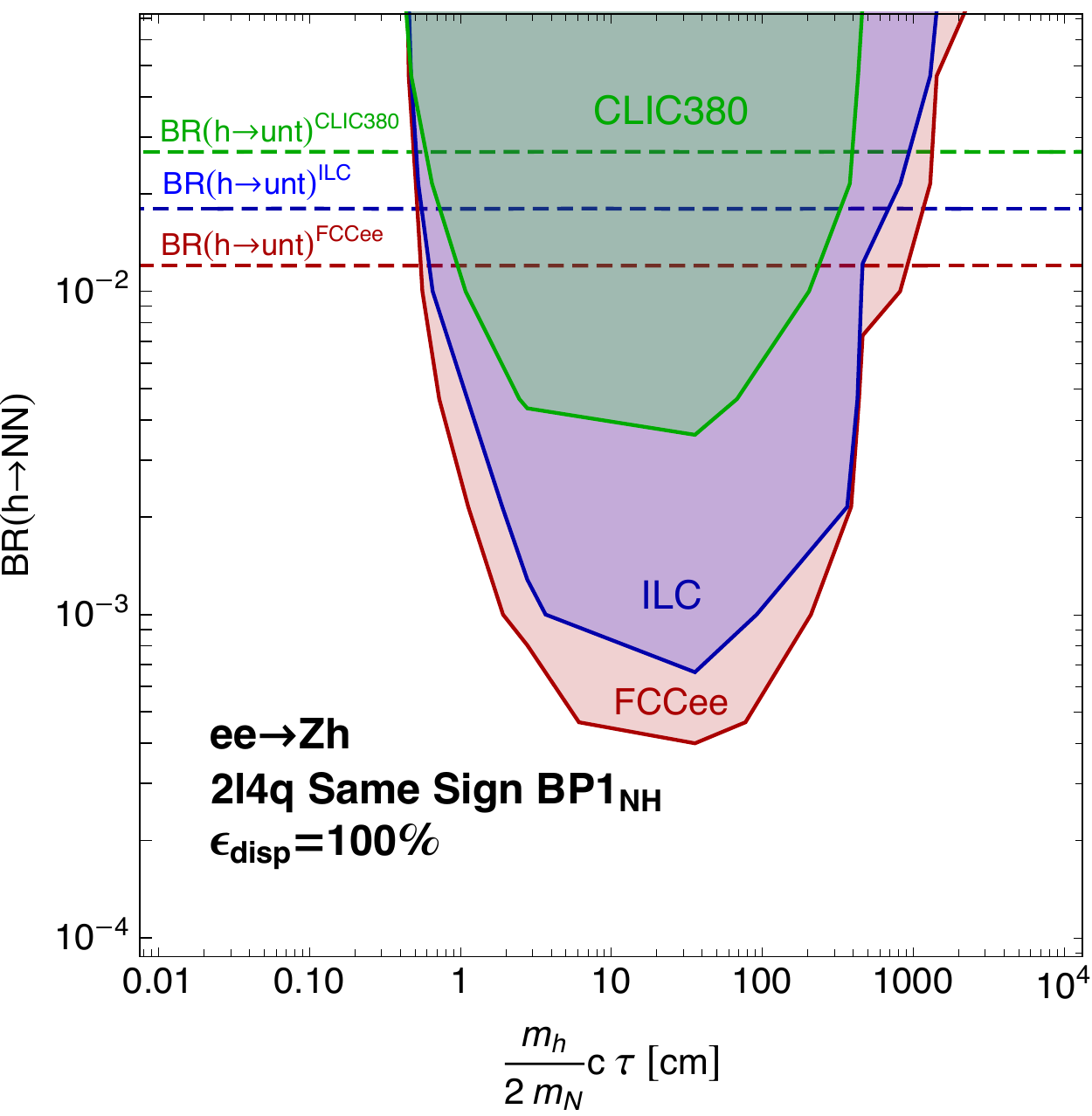}
\caption{
95\% CL exclusion for ${\bf BP1_{NH}}$ in the 
$\frac{m_h}{2 m_N}c\tau-{\rm BR}(h\to NN)$  plane from displaced decay searches in the Higgs-strahlung channel assuming $\epsilon_{\rm disp.}=50\%$ (left) and 100\% (right). The shaded red areas represent the sensitivities for the different collider options while the colored dashed lines the limits arising from untagged Higgs decay measurements.
}
\label{fig:displaced_theta2_ctauBR}
\end{center}
\end{figure}

In Fig.~\ref{fig:displaced_theta2_ctauBR} we show instead the reach of the same search projected on the plane
${\rm BR}(h\to NN)$ versus $\frac{m_h}{2 m_N}c\tau$, which is roughly the laboratory-frame decay length of the RH neutrinos. We do this again for two different choices of $\epsilon_{\rm disp.}$:  50\% (left) and 100\% (right). In the plots we also show the limits arising from untagged Higgs decays searches for the various collider options. All together we see that FCC-ee will be able to test values of the Higgs exotic BR down to 0.2\% for $\epsilon_{\rm disp.}=50\%$, largely surpassing the indirect limits from Higgs untagged decays, while ILC and CLIC-380 have a slightly weaker reach.

\subsection{Detector stable}

The last case we study is the one in which the RH neutrinos lifetime is large enough that they will decay outside the detector. In this case the decay will 
contribute to the invisible Higgs width. This quantity  can be strongly constrained at future lepton colliders, which will set a 95\% CL bound on ${\rm BR}(h\to {\rm inv.})$ of 0.22\% (FCC-ee), 0.28\% (CEPC), 0.26\% (ILC) and 0.63\%  (CLIC-380)~\cite{deBlas:2019rxi}. These limits can be directly translated on a bound on the NP scale $\Lambda$ through Eq.~\eqref{eq:GammaONH}. We obtain $\Lambda \gtrsim 360\;$TeV (FCC-ee), 320\;TeV (CEPC), $\Lambda \gtrsim 330\;$TeV (ILC) and 210\;TeV (CLIC-380) fixing $m_N=10\;$GeV. The limits degrade by roughly a factor 20\% for $m_N=35\;$GeV due to the reduced phase space. Above this mass threshold their decay will instead happen inside the detector, see  Fig.~\ref{fig:mixing_vs_ONH}.

\section{The ${\cal O}_{NB}$ operator and the s-channel $Z$ production}\label{sec:ONB}
\begin{figure}[t!]
\begin{center}
\includegraphics[width=0.48\textwidth]{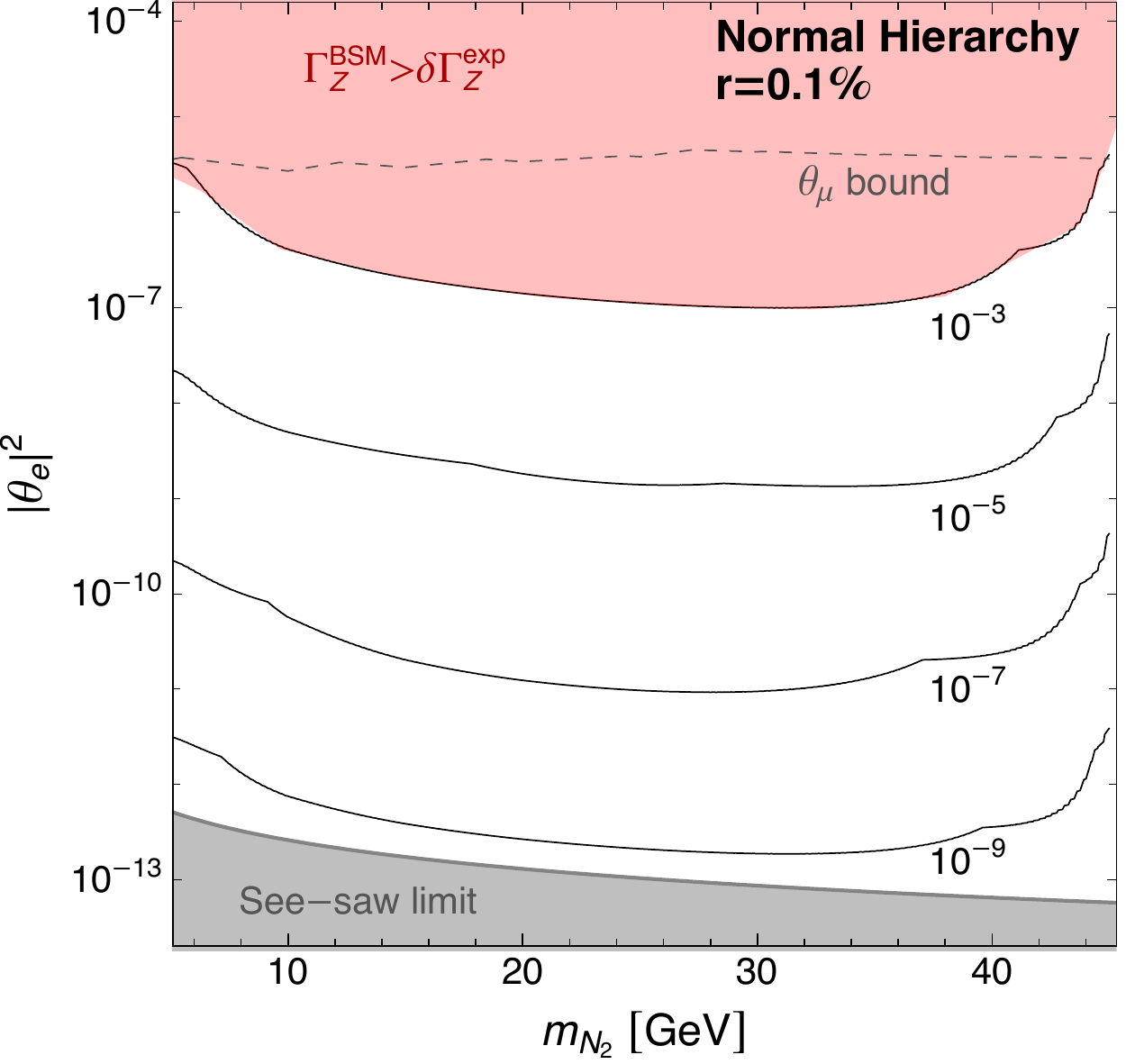}\hfill
\includegraphics[width=0.48\textwidth]{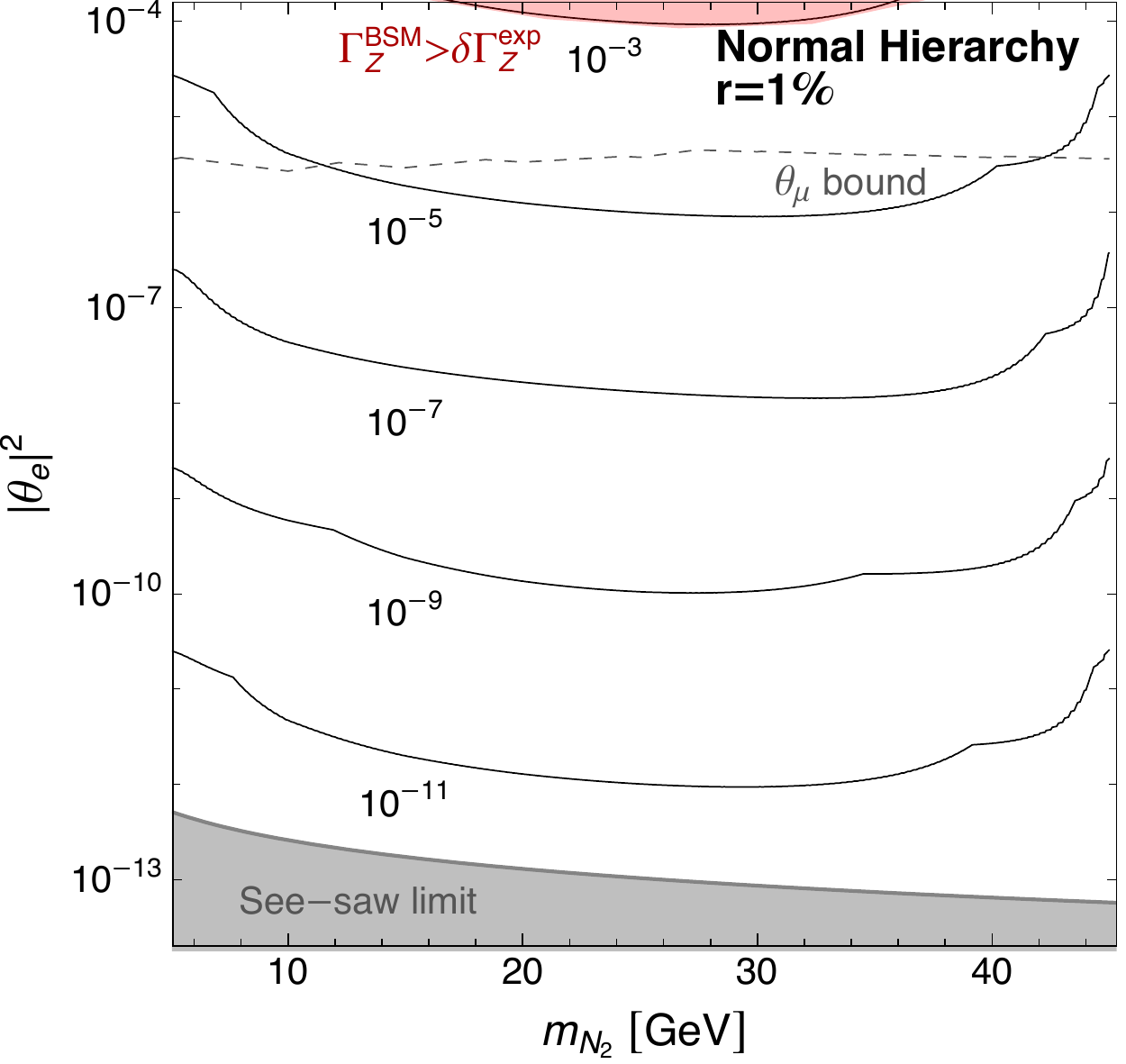}
\caption{Isocontour of equal partial widths for $N_2 \to N_1 \gamma$ and $N_2$ decaying via mixing for various choices of  ${\rm BR}(Z \to N_1 N_2)$. We fix the relative mass splitting between $N_2$ and $N_1$ to 0.1\% (left) and 1\% (right). Above the lines, for that specific ${\rm BR}(Z \to N_1 N_2)$, the decay via mixing dominates.
 In the red shaded area $\Gamma_Z^{\rm BSM}$ is larger than the current experimental uncertainty on the $Z$ boson total width.  The gray dashed line represents the limit on the mixing angle arising from existing experimental searches. In the gray shaded region  the lightness of the neutrino masses cannot be explained by the see-saw mechanism.
}
\label{fig:mixing_vs_dipole}
\end{center}
\end{figure}
We turn now to the study of the ${\cal O}_{NB}$ dipole operator. It can be generated only a loop-level by 
a scalar-fermion of vector-fermion pair with opposite hypercharges~\cite{Aparici:2009fh}.
Due to the presence of $\sigma^{\mu\nu}$, the flavor structure of this operator is antisymmetric, {\it i.e.} only different RH neutrinos can participate in the interaction. Since the operator induces a new interaction between the RH neutrinos and the SM neutral EW gauge bosons, it may provide an additional production channel for a pair of RH neutrinos through an intermediate photon or $Z$ boson. In particular, if kinematically allowed, the $Z$ boson can decay into a $N_1 N_2$ pair with a rate~\cite{Aparici:2009fh}
 \be\label{eq:ZN1N2}
\Gamma_{Z\to N_1 N_2} = \frac{2}{3\pi}\frac{|\alpha_{NB}^{12}|^2}{\Lambda^2} \frac{s_w^2}{ m_Z^3 } \lambda^{1/2}(m_Z^2,m_{N_1}^2,m_{N_2}^2) \zeta(m_Z,m_{N_1},m_{N_2}) \ ,
\ee
where $s_w$ is the sine of the Weinberg angle, $\lambda(a,b,c)=a^2+b^2+c^2-2ab-2bc-2ac$ and
\be
\zeta(m_Z,m_{N_1},m_{N_2}) = m_Z^2 (m_Z^2+m_{N_1}^2+m_{N_2}^2-6 m_{N_1} m_{N_2}\cos{2\phi_{12}})-2(m_{N_1}^2-m_{N_2}^2)^2 \ ,
\ee
with $\phi_{12}={\rm arg}[\alpha_{NB}^{12}]$. In the following we will fix $\phi_{12}=0$, {\emph{i.e}} we assume real Wilson coefficients.

Future colliders with an operating stage at $\sqrt s=m_Z$, as is the case of FCC-ee and CEPC, will produce a large number of $Z$ bosons, see Tab.~\ref{tab:colliders}, and can thus probe the operator responsible for the decay of Eq.~\eqref{eq:ZN1N2} with high precision. We focus in particular on the case of FCC-ee, where one 
expects to have $6.5\times 10^{12}$ $Z$ candidates produced, after a total integrated luminosity of $\sim 150\;$ab$^{-1}$, while CEPC might have a luminosity smaller by roughly one order of magnitude, and weaker results are generally expected. 

In addition to new production modes, the ${\cal O}_{NB}$ operator can also trigger new decay channels for the heavier RH neutrino $N_2$. In particular we have~\cite{Aparici:2009fh}
\begin{align}
\label{eq:N2N1V}
&\Gamma(N_2 \to N_1 \gamma) = \frac{2}{\pi}c_w^2 \frac{|\alpha_{NB}|^2}{\Lambda^2} m_{N_2}^3\left(1-\frac{m_{N_1}^2}{m_{N_2}^2} \right)^3 \ ,
\end{align}
while we do not consider the possibility of $N_2 \to N_1 Z$ decay which is outside of the mass range of  interest here.
Crucially, the decay mode of Eq.~\eqref{eq:N2N1V} can compete, and even dominate, with the one induced by the mixing between the active and sterile sector. It is thus important to assess in which region of parameter space the $N_2 \to N_1 \gamma$ decay rate can dominate over the decay induced by active-sterile mixing. In Fig.~\ref{fig:mixing_vs_dipole} we show such regions in the $m_{N_1}-|\theta_e|^2$ plane considering the NH case. As in previous plots, we show the limits coming from $\theta_\mu$ as gray dashed line. The black continuous lines correspond to ${\rm \Gamma}(N_2 \to N_1 \gamma) = \Gamma(N_2)_{\rm mix} $, where the latter is the total $N_2$ decay width due to mixing, {\it i.e.} in the channels listed in Tab.~\ref{tab:N_decay}, for the specific value of ${\rm BR}(Z \to N_1 N_2)$ reported. Above the lines the decays induced by mixing dominate over the decay induced by the ${\cal O}_{NB}$ operator. Since this operator involves different neutrinos, different mass splittings may give different physical situations. We quantify this by defining the relative mass splitting $r = (m_{N_2} - m_{N_1})/m_{N_2}$, which is fixed to 0.1\% in the left plot and to 1\% in the right plot. We also show in red the region in which the total BSM width $\Gamma_Z^{\rm BSM}$ is larger than the than the current experimental uncertainty on the $Z$ boson total width, $\delta \Gamma_Z^{{\rm exp}} = 2.3\;$MeV, and the region in which the see-saw mechanism is not able to reproduce the observed neutrino masses. 
As we see, in both cases for reasonable values of the branching ratio there are large regions in parameter space in which the decay via mixing dominates over $N_2 \to N_1 \gamma$.

As a consequence, in this Section we will focus on the signatures of $N_2$ decaying via mixing. Notice that, barring the contribution of $d=6$ operators on which we will comment in Sec.~\ref{sec:d6}, this is always true also for $N_1$.  As it happened in Sec.~\ref{sec:ONH}, we find that for mixing angles compatible with experimental bounds, the $N_1 N_2$ production via $Z$ decay can easily dominate with respect to the $N\nu$ production via mixing. In the following Sections we will thus analyze the sensitivity of future experiments for the case in which the $N_1N_2$ production occurs via $Z$ decay and $N_{1,2}$ decay via mixing. As already discussed, also in this case the decay can be prompt, displaced or outside the detector. For simplicity we fix $m_{N_2}-m_{N_1}\ll m_{N_{1,2}}$, {\emph{i.e.}} we study the case of almost degenerate RH neutrinos.

\subsection{Prompt decay}

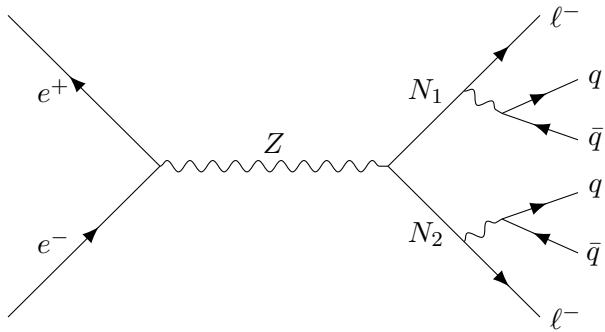
\begin{figure}
\begin{center}
\adjustbox{valign=m}{
 \begin{tikzpicture}
\draw[f] (-2,-2) -- (0,0) node[midway, xshift=-4mm]{$e^-$};
\draw[f] (0,0) -- (-2,2) node[midway, xshift=-4mm]{$e^+$};
\draw[v] (0,0) -- (3,0) node[midway, yshift=3mm]{$Z$};
\draw (3,0) -- (4, 1) node[midway,yshift=5mm]{$N_1$};
\draw (3,0) -- (4, -1) node[midway,yshift=-4mm]{$N_2$};
\draw[f] (4,1) -- (5, 2) node[right]{$\ell^-$};
\draw[f] (4,-1) -- (5, -2) node[right]{$\ell^-$};
\draw[v] (4,1) -- (4.5, 0.7);
\draw[v] (4,-1) -- (4.5, -0.7);
\draw[f] (4.5, 0.7) -- (5.5, 1.15) node[right]{$q$};
\draw[f] (5.5, 0.35) node[right]{$\bar q$}-- (4.5, 0.7);
\draw[f] (5.5, -1.15) -- (4.5, -0.7) node[midway,xshift=7mm,yshift=-2.5mm]{$ \bar q$};
\draw[f] (4.5, -0.7) node[midway,xshift=57.3mm,yshift=-3mm]{$q$}-- (5.5, -0.35);
\end{tikzpicture}} 
\end{center}
\caption{Feynman diagram for $2\ell 4q$ production in the SS leptons final state through s-channel $Z$ production. }\label{fig:feynman_ONB}
\end{figure}

\begin{figure}[t!]
\begin{center}
\includegraphics[width=0.48\textwidth]{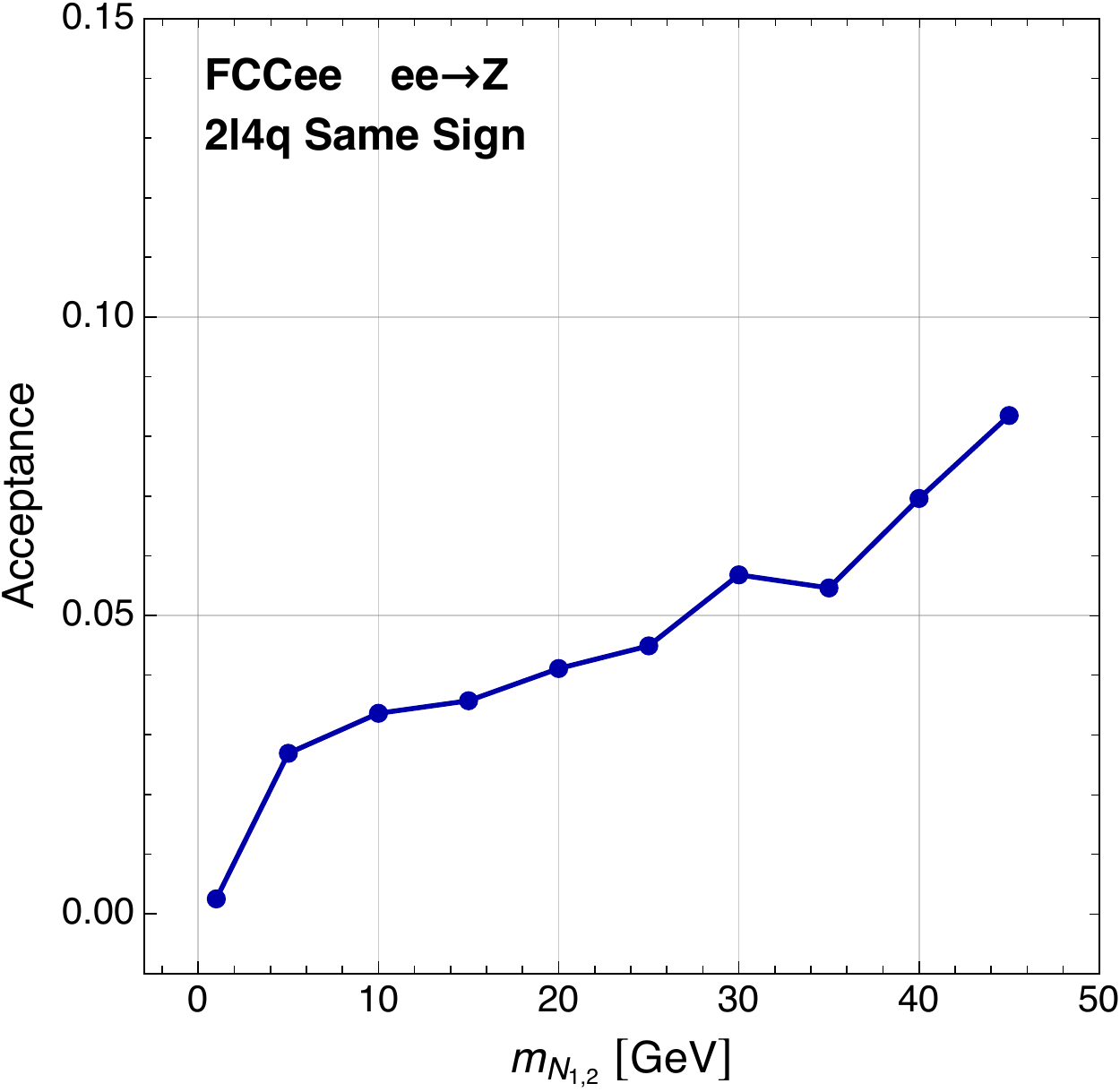}
\caption{\small Parton level acceptances for the $2\ell 4q$ final state in the $Z$ channel topology with a leptonically decaying $Z$ boson and with the selection cut described in the main text.}
\label{fig:acceptance_ONB}
\end{center}
\end{figure}

We again focus on the final state with the highest rate, {\emph{i.e.}} the $2\ell 4q$ channel, restricting our analysis to the final state with a pair of SS leptons. This process is shown in Fig.~\ref{fig:feynman_ONB}. 
We require exactly two SS leptons with $p_T>2.5\;$GeV and $|\eta|<2.44$, while jets should satisfy $p_T>5\;$GeV and $|\eta| <2.4$. Leptons are required to be separated by $\Delta R>0.15$ among themselves and with respect to the selected jets.  With these selections, the parton level acceptances for the $e^+ e^+$ flavor combination are reported in Fig.~\ref{fig:acceptance_ONB}. 

Regarding the background analysis, in the case of the ${\cal O}_{NH}$ operator we assumed 
that the physical background mimicking 
the Higgs-stralung production and  subsequent six-body decay
in a sample of around $10^6$ Higgs bosons would be well under control and negligible in first approximation.  
On the other hand, in  case of the FCC-ee  production of  ${\cal O}(10^{12})$ $Z$ bosons at the $Z$ peak,  quite a number of reducible backgrounds
are expected to limit the statistical reach of the $Z$ boson sample, depending
both on the limited detector performances and on the collider characteristics.
We then expect that a realistic background analysis for the 
$Z\to \ell^+ \ell^+ 4q$ channel
might degrade the ideal background free estimate in a non negligible way.

As an example of how different reducible backgrounds can degrade the  $Z$ sample sensitivity, we will consider below the possible background arising
from the limited lepton-charge identification power of a  LHC-like detector. 
 Indeed,
conservation of lepton number implies that no genuine  irreducible  
$Z\to \ell^+ \ell^+ 4q$  background arises in the SM. As regarding the $Z\to\ell^+ \ell^+ 2\nu 4q$ background, 
for which we expect the cross-section to be relatively small,  this might also be efficiently reduced by asking for limited missing  energy in the events. 
On the other hand, a realistic analysis of the reducible  SM background  $Z\to \ell^+ \ell^- 4q$, where a lepton charge is misidentified, or  an hadron is misidentified as a lepton, would require a dedicated full simulation at the experimental level, which is beyond the scope of the current work. For the lepton-charge misidentification effect, 
assuming LHC performances of lepton charge identification, 
we provide an approximate estimate of the corresponding background by applying 
a (flat) mis-identification probability factor of  $\epsilon^\ell_{{\rm misID}}=10^{-3}$~\cite{Aaboud:2019ynx} to the partonic cross-section $e^+ e^- 4q$ and $\mu^+ \mu^-4q$ after the signal selections described in the text~\footnote{Note that at the LHC the probability of mis-identifying the charge of a muon is generally negligible, since muon tracks are measured both in the inner detector and in the muon spectrometer.}. With this procedure we obtain a background yield of $\sigma_{\ell^+ \ell^- 4q}\times 2 \times \epsilon^\ell_{{\rm misID}}(1-\epsilon^\ell_{{\rm misID}})\simeq 130\;$fb$\times \epsilon^\ell_{{\rm misID}}(1-\epsilon^\ell_{{\rm misID}})\simeq 0.26\;$fb, which we use as estimate for our analysis. 

\begin{figure}[t!]
\begin{center}
\includegraphics[width=0.48\textwidth]{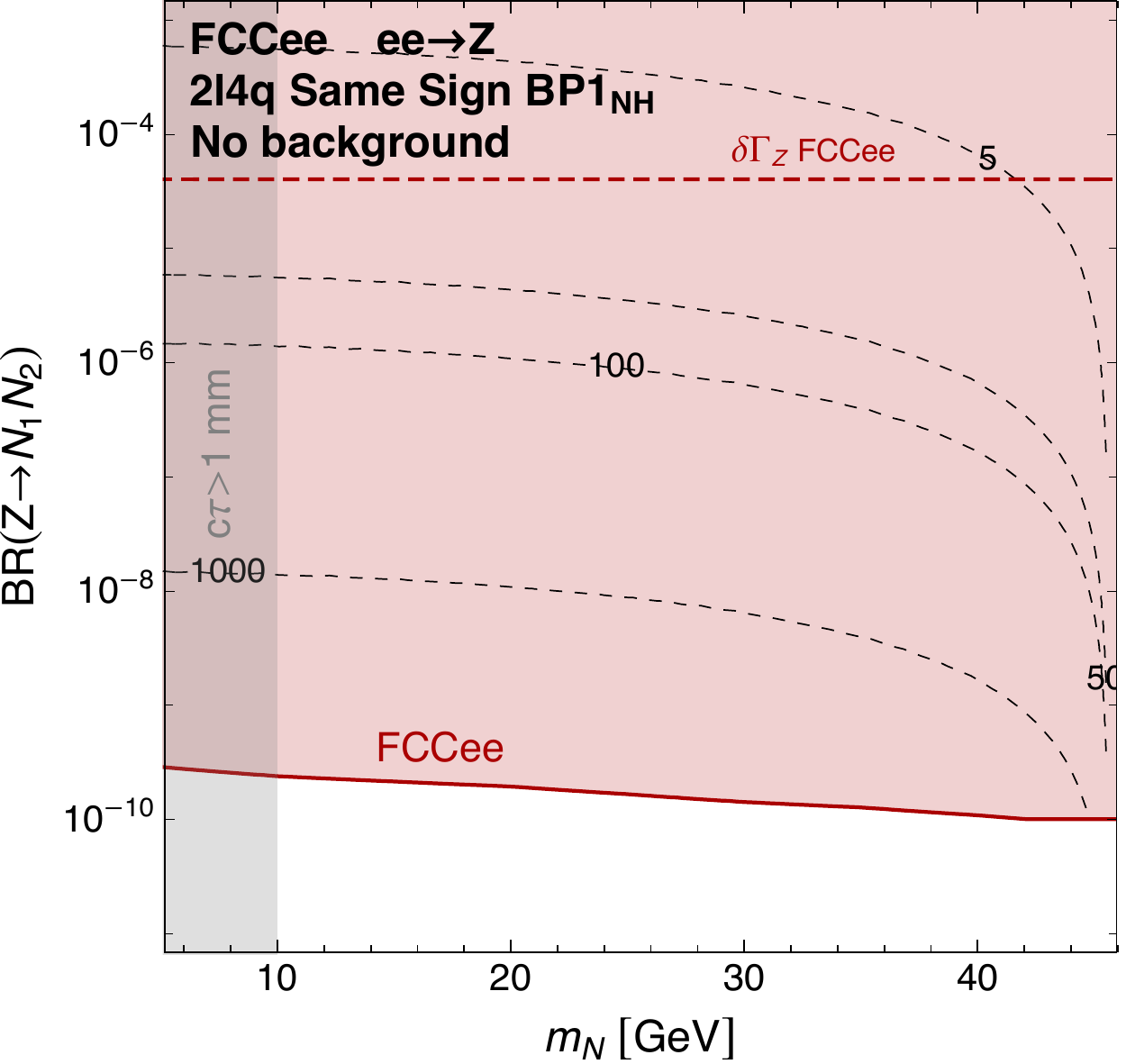}\hfill
\includegraphics[width=0.48\textwidth]{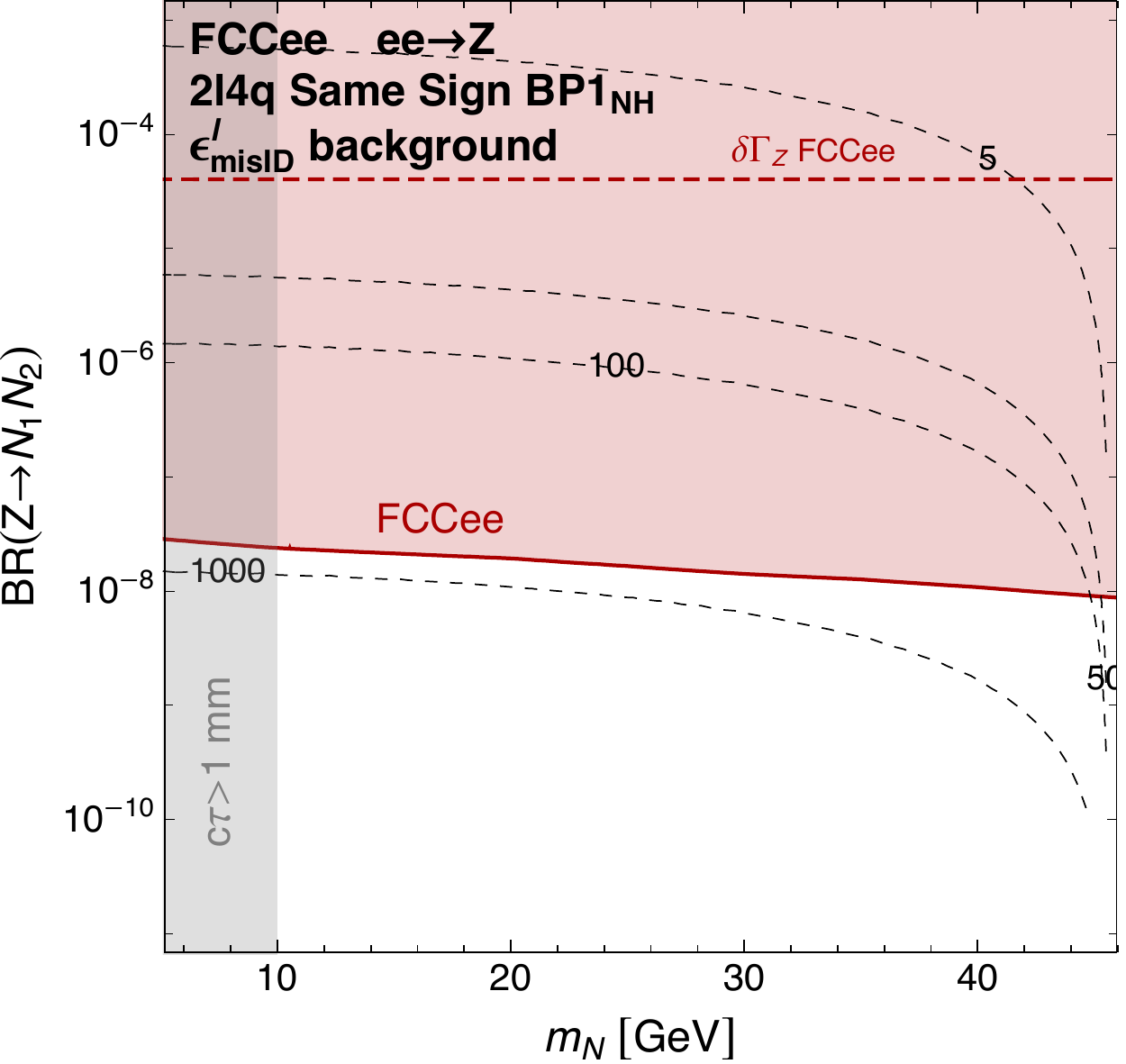}
\caption{\small 95\% CL exclusion for ${\bf BP1_{NH}}$
 from prompt searches in the s-channel $Z$ production channel.
 The red shaded area represent the exclusion for the case of FCC-ee. Also shown as a red dashed line the limits arising from the measurement of the $Z$ boson total width and  as gray dashed lines the isocontours of NP scale $\Lambda$ in TeV. In the gray shaded area $c\tau >1\;$mm and the RH neutrino cannot decay promptly without being excluded by experimental searches. In the left plot we assume zero background while in the right plot we include the SM reducible background arising from lepton charge mis-identification.}
\label{fig:reach_runz_prompt}
\end{center}
\end{figure}

We show our results for ${\bf BP1_{NH}}$ in Fig.~\ref{fig:reach_runz_prompt} in the $m_N-{\rm BR}(Z\to N_1 N2)$ plane.  As for the ${\cal O}_{NH}$ case, in computing our limits we are summing 
on all the lepton flavor combinations from the $N_1 N_2$ decay and we are considering all lepton charges configurations that give rise to a SS lepton final state. 
In the left panel we assume zero background, while in the right panel we estimate the background coming from lepton charge mis-identification following the procedure outline above. The red region is the one that will be probed by the FCC-ee, while the horizontal dashed red line represent the expected sensitivity on the $Z$ boson width of $100\;$keV~\cite{Gomez-Ceballos:2013zzn}. From the figure it is clear that, when considering the presence of the lepton charge mis-identification background, FCC-ee will be able to exclude values of ${\rm BR}(Z\to N_1 N_2)$ down to roughly $10^{-8}$. This corresponds to a reach on $\Lambda \simeq 10^3\;$TeV. Given that this operator can only be generated at loop-level, this bounds is rescaled by a factor $16\pi^2$ when mapped into the physical masses and couplings of an ultraviolet complete model.
We stress that the present estimate might be further degraded by other background sources, such as other particle identification inefficiencies which will be strongly dependent on the actual detector performance.

\subsection{Displaced decay}

\begin{figure}[t!]
\begin{center}
\includegraphics[width=0.48\textwidth]{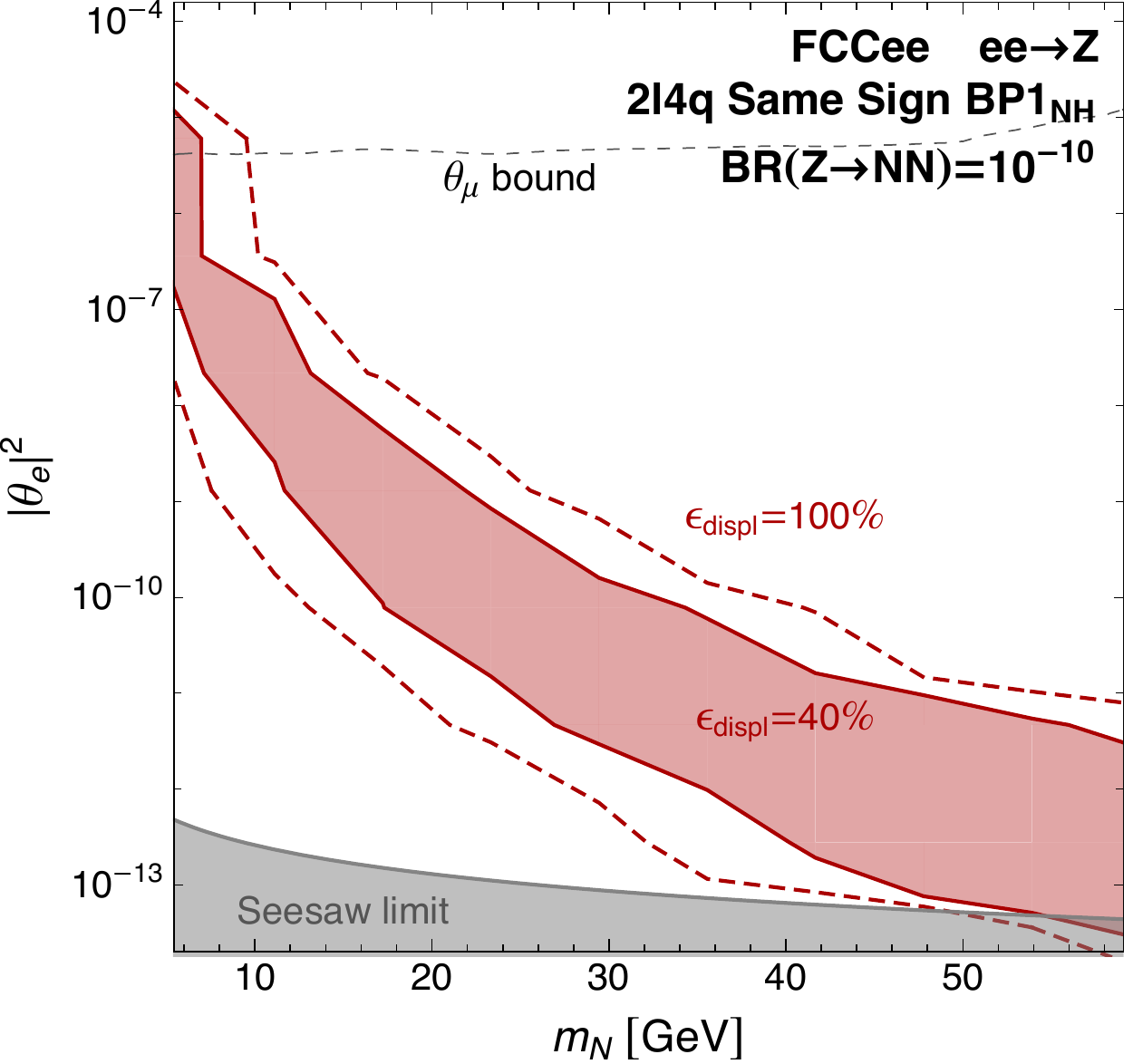}\hfill
\includegraphics[width=0.47\textwidth]{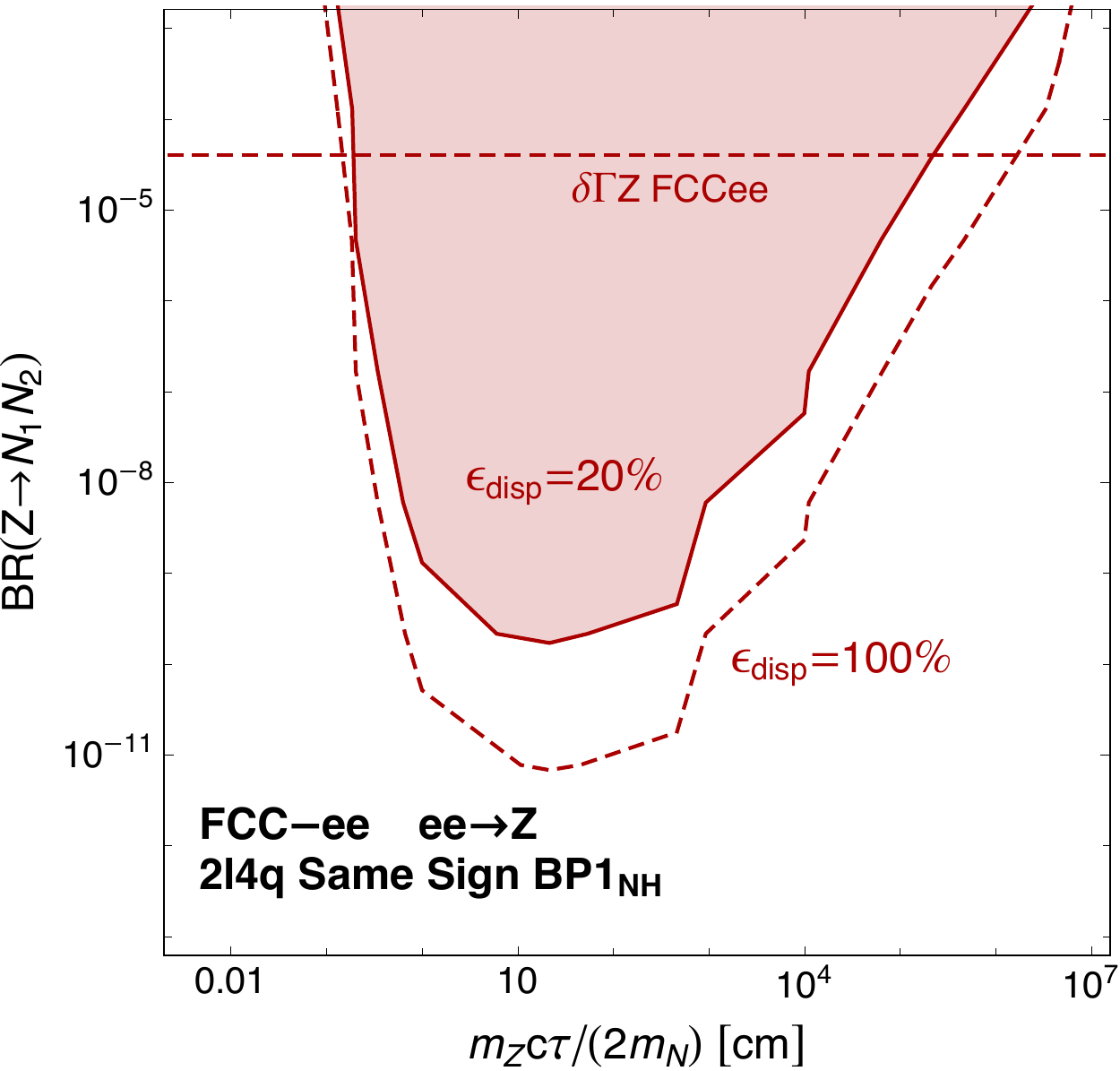}
\caption{{\emph{Left:}}
95\% CL exclusion  for ${\bf BP1_{NH}}$ in the $m_N-|\theta_e|^2$ plane from displaced decay searches assuming ${\rm BR}(Z\to N_1N_2)=10^{-10}$ for the case of FCC-ee. The gray dashed line represents the limit on the mixing angle arising from existing experimental searches. In the gray shaded region  the lightness of the neutrino masses cannot be explained by the see-saw mechanism.
{\emph{Right}}: 
95\% CL exclusion for ${\bf BP1_{NH}}$ in the 
$\frac{m_Z}{2 m_N}c\tau-{\rm BR}(h\to N_1N_2)$  plane from displaced decay searches. The horizontal red dashed line represents the limit arising from the FCC-ee $Z$ boson total width measurement with an uncertainty of $100\;$keV.
}
\label{fig:reach_runz_displaced}
\end{center}
\end{figure}

For the case in which the RH neutrinos decay displaced from the primary vertex we follow again the strategy outlined in Sec.~\ref{sec:ONH-disp}. As opposed to the case of prompt decay, in this case we expect negligible irreducible SM background.
We show in the left panel of Fig.~\ref{fig:reach_runz_displaced} the FCC-ee reach for the ${\bf BP1_{IH}}$ benchmark,  projected in the $m_N-|\theta_e|^2$. We fix ${\rm BR}(Z\to N_1N_2)=10^{-10}$ and assume different efficiencies $\epsilon_{\rm displ}$, as reported in the plots.  The gray shaded are represents the see-saw limit, while the values of the mixing angles above the gray dashed line are excluded by experimental searches. As for the case of the ${\cal O}_{NH}$ operator our results show that the search for RH neutrinos arising from $Z$ decay and decaying displaced from the primary vertex can offer a great handle in testing the active-sterile mixing angle. Again this is due to the fact that the production cross-section does not depend on this quantity, but only on the $Z$ decay rate into the NP final state. In the right panel of the same Figure we show instead the FCC-ee reach projected in the $\frac{m_Z}{2 m_N}c\tau-{\rm BR}(h\to N_1N_2)$ plane. We do this again for two different choices of $\epsilon_{\rm displ}$ as reported in the plot. All together we see that FCC-ee will be able to test values of  the $Z$ boson exotic branching ratio down to $10^{-9}$ for $\epsilon_{\rm disp.}=20\%$, largely surpassing
the limit arising from the FCC-ee $Z$ boson total width measurement with an uncertainty of $100\;$keV, which is represented by the horizontal red dashed line.

\subsection{Detector stable}

The last case we study is again the possibility that the RH neutrinos lifetime is large enough to cause them to decay outside the detector. Similarly to the ${\cal O}_{NH}$ operator, in this case the decay will 
contribute to the invisible $Z$ width. FCC-ee will measure the ratio $R_\nu = \Gamma_{Z\to {\rm inv}} /\Gamma_{Z\to \ell\ell}$ at the level of $0.27\times 10^{-3}$~\cite{deBlas:2019rxi} which, under the SM hypothesis, correspond to an additional contribution to invisible decay width of the $Z$ boson smaller than $\sim 135\;$KeV. This limit can be translated in a bound on the NP scale $\Lambda$ which will be constrained by this measurement to be $\Lambda \gtrsim 16\;$TeV for $m_N\sim10\;$GeV while this limits degrade down to $\Lambda \gtrsim 9\;$TeV for $m_{N}=35\;$GeV due to phase space effect. Above this mass threshold their decay will happen inside the detector, see  Fig.~\ref{fig:mixing_vs_ONH}.


\section{The impact of $d=6$ operators}\label{sec:d6}

We have so far considered the phenomenology
induced by the presence of $d=5$ operators in the $\nu$SMEFT, discussing how they can induce additional RH neutrinos production and decay modes and showing how they can be efficiently tested at future Higgs Factories. It is important however to notice that at $d=6$ many more operators are present, and they might give observable signatures. For example, among the operators involving the Higgs field, of particular interest are 
\begin{align}\label{eq:d=6-A}
& \mathcal{O}_{LNB} =\alpha_{LNB} (\bar{L}\sigma^{\mu\nu}N)B_{\mu\nu}\tilde{H}  \ , \nn \\
& \mathcal{O}_{LNW} = \alpha_{LNW} (\bar{L}\sigma^{\mu\nu}N)W_{\mu\nu}\tilde{H} \ .
\end{align}
They will trigger the decay to an active neutrino and a photon with a rate
\begin{equation}
\Gamma(N \rightarrow \nu_i \gamma) \simeq \frac{m_N^3v^2}{4\pi \Lambda^4}\Big(\alpha_{LNB} + s_w \alpha_{LNW} \Big)^2,
\end{equation}
where, with an abuse of notation, we also indicate with $\alpha_{LNB}$ and $\alpha_{LNW}$ the relevant entries of the corresponding Wilson coefficient matrices.  
 Also this decay mode can in principle dominate over the one induced via the active-sterile mixing in some regions of the parameter space. A detailed analysis would proceed in a very similar fashion to the one described in Sec.~\ref{sec:ONB} for the dipole operator $\mathcal{O}_{NB}$, with however an additional suppression due to their higher dimensionality. 
These two operators have also been recently studied in the context of LHC in ref.~\cite{Butterworth:2019iff} where a bound on $\Lambda \gtrsim 2.2\;$TeV has been obtained via the $p p \to h \to \nu N \gamma$ with subsequent decay $N\to \nu \gamma$. Notice that in deriving this bound the Authors of~\cite{Butterworth:2019iff} have considered  also the Higgs decay to be triggered by the $d=6$ operators of Eq.~\eqref{eq:d=6-A}.

At $d=6$ it is also interesting to notice the presence of four fermions operators. In the case of future Higgs Factories of particular relevance is
\be
\mathcal{O}_{Ne} = \alpha_{NE} (\bar{N}\gamma^{\mu}N)(\bar{e}_R\gamma_{\mu}e_R) \ , 
\ee
which triggers a direct production channel for the RH neutrinos pair with a rate
\be
\sigma(ee\to NN) \simeq \left(\frac{\sqrt s }{8 \pi \Lambda^2}\alpha_{NE}\right)^2  \beta \left(1+\frac{\beta^3}{3}\right) \ , 
\ee
where $\beta = \sqrt{1-\frac{4 m_N^2}{s}}$, which gives
\be
 \sigma(ee\to NN)\simeq 1\;{\rm fb} \left( \frac{2.5\;{\rm TeV}}{\Lambda}\right)^4\left(\frac{\sqrt s}{240\;{\rm GeV}}\right)^2
 \ee
for $m_N=10\;$GeV. However, unless the coefficients of the $d=5$ operator ${\cal O}_{NH}$ and the one of the $d=6$  operator ${\cal O}_{Ne}$ have a different scaling, for equal NP scale $\Lambda$ the production induced by the former is always dominant with respect to the latter for regions where the additional Higgs decay width is compatible with current constraints, $\Lambda \gtrsim 50\;$TeV.
 The situation can be drastically different in the case where the underlying theory has a particular symmetry, as in the case of Minimal Flavor violation recently analyzed in~\cite{Barducci:2020ncz}. In this case the Higgs operator receives an extra suppression of a factor $\sim m_N/ \Lambda$, while four fermions operators as ${\cal O}_{Ne}$ will not be affected by the insertion of any spurion.
In this case the main production mechanism, especially for low neutrino masses and large center of mass energies, will be via the one through the $d=6$ operators.

\section{Conclusions}\label{sec:conc}

The observed pattern of neutrino masses and oscillation parameters require extending the Standard Model. One of  the simplest possibilities is to add to the SM particle content two or more RH neutrinos. In this framework active neutrino masses compatible with current experimental measurements are generated via the see-saw mechanism, through an interplay of the active-sterile Yukawa coupling and the RH neutrinos Majorana mass. Naturalness consideration and the observation of a large baryon asymmetry in the Universe, motivates the study of scenarios where RH neutrinos have a mass $M_N$ at around the EW scale $v$, hence testable at current and future collider experiments. In general, such RH neutrino states are assumed to be produced through mixing with the SM sector, which is also responsible for their decay. The presence of additional New Physics at a scale $\Lambda \gg v, M_N$ can drastically modify their phenomenology and have thus a huge impact for present and future experimental search strategies. The impact of these extra deformations can be parametrized at low energy as an effective field theory with $d>4$ operators built out from SM and RH neutrino fields. In this work we have focused on the effect of the two new $d=5$ operators containing both the SM and the RH neutrinos, ${\cal O}_{NH}$ and ${\cal O}_{NB}$. They induce additional production modes for RH neutrino pairs through the decay of the Higgs and the $Z$ boson respectively. 
We have then studied the phenomenology of these two operators at future Higgs Factories, such as FCC-ee, CEPC, ILC and CLIC-380, for different regimes of RH neutrinos lifetimes. In particular we have considered RH neutrinos with prompt, displaced or outside the detector decays. For both operators we have shown that,
in favorable scenarios for detector performance and systematics, future Higgs Factories have a great potential in testing the NP scale $\Lambda$, well above the limit that can be set by indirect probes such as the search for additional untagged Higgs decay or the measurement of the $Z$ boson decay width. In particular, in the case of prompt decays the Higgs  and $Z$ branching ratios into $NN$ pairs can be tested at the level of $10^{-3}$ and $10^{-8}$, respectively, in the most favorable scenarios, while search for displaced decays can probe active-sterile mixing angles down to the see-saw limit of $|\theta_e|^2\sim 10^{-13}$. We have moreover discussed the possibility of disentangling the underlying flavor structure should a signal be observed in future experiments, and commented on possible additional signatures that can be triggered by $d=6$ operators, and that deserve a separate and dedicated study.

\section*{Acknowledgments}
We thank Roberto Franceschini for comments on the draft.
EB acknowledges financial support from FAPESP under contracts 2015/25884-4 and 2019/15149-6 and is indebted to the Theoretical Particle Physics and Cosmology group at King's College London for its hospitality. AC acknowledges hospitality from the MPP of Munich.
AC acknowledges support from the Generalitat Valenciana (GVA) through the GenT program (CIDEGENT/2018/019).
 PH acknowledge support from the GVA project PROMETEO/2019/083, as well
as the national grant FPA2017-85985-P and  the European grant H2020-MSCA-ITN-
2019//860881-HIDDeN. 

\newpage
\bibliographystyle{JHEP}
{\footnotesize
\bibliography{biblio}}

\end{document}